\documentclass[article,twocolumn,nofootinbib,superscriptaddress]{revtex4}
\input epsf
\usepackage{graphics}
\usepackage{amsmath}
\usepackage{amssymb}
\usepackage{bm}

\usepackage{color} 
\usepackage{dcolumn}
\usepackage{hyphenat}
\usepackage{hyperref}
\usepackage[dvipsnames]{xcolor}
\usepackage{booktabs}
\usepackage[normalem]{ulem}

\hypersetup{
    colorlinks=true,
    linkcolor=red,
    citecolor=blue,
}

\def\be{\begin{equation}}
\def\ee{\end{equation}}
\def\ba{\begin{eqnarray}}
\def\ea{\end{eqnarray}}

\usepackage{graphicx}

\begin{document}

\title{New MGCAMB tests of gravity with CosmoMC and Cobaya}

\author{Zhuangfei Wang}
\affiliation{Department of Physics, Simon Fraser University, Burnaby, BC, V5A 1S6, Canada}
\author{Seyed Hamidreza Mirpoorian}
\affiliation{Department of Physics, Simon Fraser University, Burnaby, BC, V5A 1S6, Canada}
\author{Levon Pogosian} 
\affiliation{Department of Physics, Simon Fraser University, Burnaby, BC, V5A 1S6, Canada}
\author{Alessandra Silvestri}
\affiliation{Institute Lorentz, Leiden University, PO Box 9506, Leiden 2300 RA, The Netherlands}
\author{Gong-Bo Zhao} 
\affiliation{National Astronomy Observatories, Chinese Academy of Science, Beijing, 100101, P.R.China}
\affiliation{University of Chinese Academy of Sciences, Beijing, 100049, P.R.China}
\affiliation{Institute for Frontiers in Astronomy and Astrophysics, Beijing Normal University, Beijing, 102206, P.R.China}

\begin{abstract}

We present a new version of MGCAMB, a patch for the Einstein-Boltzmann solver CAMB for cosmological tests of gravity.  New features include a new cubic-spline parameterization allowing for a simultaneous reconstruction of $\mu$, $\Sigma$ and the dark energy density fraction $\Omega_X$ as functions of redshift, the option to work with a direct implementation of $\mu$, $\Sigma$ (instead of converting to $\mu$, $\gamma$ first), along with the option to test models with a scalar field coupled only to dark matter, and the option to include dark energy perturbations when working with $w\ne -1$ backgrounds, to restore consistency with CAMB in the GR limit. This version of MGCAMB comes with a Python wrapper to run it directly from the Python interface, an implementation in the latest version of CosmoMC, and can be used with Cobaya. 
\end{abstract}  

\maketitle

\section{Introduction}

The discovery of cosmic acceleration~\cite{Perlmutter:1998np,Riess:1998cb}, along with the unnatural smallness of the implied cosmological constant $\Lambda$~\cite{Weinberg:1988cp,Burgess:2013ara} and the unexplained nature of the cold dark matter (CDM), stimulated extensive studies of alternative gravity theories and their phenomenological signatures~\cite{Silvestri:2009hh,Clifton:2011jh,Joyce:2014kja,Koyama:2015vza,Ishak:2018his}. Further interest is driven by the opportunities for testing General Relativity afforded by the current and future cosmological surveys and, more recently, the emergence of tensions between datasets \cite{Abdalla:2022yfr} within the $\Lambda$CDM model. Modified Growth with {\tt CAMB} ({\tt MGCAMB}), first introduced in 2008~\cite{Zhao:2008bn} and significantly upgraded in 2011~\cite{Hojjati:2011ix} and 2019~\cite{Zucca:2019xhg}, is a patch for the popular Einstein-Boltzmann solver {\tt CAMB}~\cite{Lewis:1999bs,camb}, allowing one to compute cosmological observables in models with modified relations between the gravitational metric potentials and the matter density contrast. It has been used in conjunction with Monte-Carlo Markov Chain samplers, such as {\tt CosmoMC}~\cite{Lewis:2002ah,cosmomc}, to constrain modifications of gravity on cosmological scales~\cite{Simpson:2012ra,Ade:2015rim,DES:2018ufa}. 

The field of cosmological tests of gravity has matured significantly over the past couple of decades. Studies evolved from the phenomenology of specific models, such as $f(R)$~\cite{Capozziello:2003tk,Carroll:2003wy,Hu:2007nk,Pogosian:2007sw} and DGP~\cite{Dvali:2000rv,Koyama:2005kd}, to the development of general frameworks~\cite{Gubitosi:2012hu,Bloomfield:2012ff,Gleyzes:2013ooa,Bloomfield:2013efa,Silvestri:2013ne,Bellini:2014fua,Gleyzes:2014rba,Gleyzes:2015pma} for studying broad classes of modified gravity theories, such as Horndeski~\cite{Horndeski:1974wa,Deffayet:2011gz} and beyond~\cite{Gleyzes:2014qga,Gleyzes:2014dya,Langlois:2015cwa,BenAchour:2016cay}, along with the numerical tools for interpreting observations within these frameworks, such as {\tt EFTCAMB}~\cite{Hu:2013twa,2014arXiv1405.3590H,Raveri:2014cka} and {\tt hi\_class}~\cite{Zumalacarregui:2016pph,Bellini:2019syt}. The essential difference between {\tt MGCAMB} (and similar software, like {\tt MGCLASS}~\cite{Sakr:2021ylx} and {\tt ISiTGR}~\cite{Dossett:2011tn,Garcia-Quintero:2019xal,Garcia-Quintero:2020mja}) and codes like {\tt EFTCAMB} and {\tt hi\_class} is that the latter are exact tools for testing scalar-tensor theories, albeit of most general type, while the former are purely phenomenological, helping constrain departures from GR that are more directly probed by large-scale structure surveys. In practice, one always has to choose a particular parameterization and such choices are often driven by intuition gained from scalar-tensor theories. Thus, there is a good degree of synergy between Horndeski-based tools and {\tt MGCAMB} when it comes to obtaining theoretical priors~\cite{Espejo:2018hxa} on the phenomenological variables and the interpretation of the constraints~\cite{Pogosian:2016pwr}. A recent example of such a synergistic approach was the joint reconstruction~\cite{Pogosian:2021mcs,Raveri:2021dbu} of the two {\tt MGCAMB} functions, $\mu$ and $\Sigma$, parameterizing the relations between the matter density contrast and the Newtonian and Weyl potentials, respectively, and the dark energy (DE) density fraction, $\Omega_X$, with the help of a Horndeski correlation prior obtained using {\tt EFTCAMB}.

This release of {\tt MGCAMB} is publicly available at \url{https://github.com/sfu-cosmo/MGCAMB} and comes with tools for using it with {\tt Cobaya}~\cite{Lewis:2002ah,Torrado:2020dgo,cobaya} and an implementation in the latest version of {\tt CosmoMC} ~\cite{Lewis:2002ah,Lewis:2013hha,cosmomc}. The added features include new models, such as a scalar field coupled only to dark matter in the quasi-static approximation, the option to include DE perturbations when working with $w\ne -1$ backgrounds, and a new binned parameterization allowing for a simultaneous reconstruction of $\mu$, $\Sigma$ and $\Omega_X$ as functions of redshift. In what follows, we describe the new features in detail and demonstrate their use in a few representative examples.
 
\section{Overview of MGCAMB and its new features} 
\label{sec2}

Let us start  briefly reviewing the framework of {\tt MGCAMB} and the options available in the 2019 version, followed by a detailed description of the new features introduced in the current release. A more comprehensive description of {\tt MGCAMB} is provided in~\cite{Zucca:2019xhg}.

\subsection{MGCAMB and its 2019 version} 
\label{sec2A}

In {\tt MGCAMB}, departures from GR are encoded in two phenomenological functions of the scale factor $a$ and Fourier number $k$, $\mu(a,k)$ and $\gamma(a,k)$, introduced in the Newtonian gauge linearized Einstein equations as
\ba
k^2 \Psi=-4 \pi G \ \mu(a, k) \ a^2 [\rho \Delta+3(\rho+P) \sigma ]
\label{eq:Poisson1} \\
k^2[\Phi-\gamma(a, k) \Psi]=12 \pi G \ \mu(a, k) \ a^2 (\rho+P) \sigma \ ,
\label{eq:Poisson2}
\ea
where $\rho \Delta \equiv \sum_i \rho_i \Delta_i$ and $(\rho+P) \sigma \equiv \sum_i (\rho_i+P_i) \sigma_i$, $\Delta_i$ is the comoving energy density contrast, $\rho_i$ and $P_i$ are the background density and pressure, and $\sigma_i
$ is the anisotropic stress of individual particle fluids labeled by $i \in \{b,c,\gamma,\nu\}$ for baryons, CDM, photons and neutrinos, respectively, and $\Psi$ and $\Phi$ are the metric potentials, defined via
\be
d s^{2}=a^{2}(\tau)\left[-(1+2 \Psi) d \tau^{2}+(1-2 \Phi) d x^{i} d x_{i}\right] \ ,
\label{eq:separation}
\ee
where $\tau$ denotes the conformal time. At late times, when the anisotropic stress contribution from photons and neutrinos is negligible, the equations become 
\ba
k^2 \Psi &=& -4 \pi G \mu(a, k) a^2 \rho \Delta
\label{eq:Poisson1-late} \\
\Phi &=& \gamma(a, k) \Psi
\label{eq:Poisson2-late} \ ,
\ea
A popular and equivalent parameterization employs $\Sigma(a,k)$, instead of $\gamma(a,k)$, defined as 
\be
k^{2}(\Phi+\Psi)=-4 \pi G \ \Sigma(a,k) \ a^{2} [2\rho \Delta+3(\rho+P) \sigma] \ .
\label{def:Sigma}
\ee
In the limit of negligible photon and neutrino anisotropic stress, {\it i.e.} well after the onset of matter domination, $\Sigma$ and $\gamma$ are simply related via 
\be
\Sigma = {\mu \over 2}(1+\gamma) .
\label{eq:mu-gamma-Sigma}
\ee
In the 2019 version of {\tt MGCAMB}, when the user opted to specify $\mu$ and $\Sigma$, they would be converted into $\mu$ and $\gamma$ using (\ref{eq:mu-gamma-Sigma}), with the subsequent calculations carried by the code in terms of the latter. One of the new features in the 2023 version is the direct implementation of the $\mu$ and $\Sigma$ parameterization in the equations, eliminating the need to assume validity of (\ref{eq:mu-gamma-Sigma}), {\it i.e.} the need to neglect the anisotropic stress from relativistic particle species.

We note that variations of the gravitational coupling in the Solar System are tightly constrained by lunar ranging and other experiments. The parameterization in Eq.~(\ref{eq:Poisson1-late}) implies an effective coupling $G_{\rm eff} \equiv \mu(a,k) G$, determining the gravitational clustering of matter on cosmological scales, where $G$ is the Newton's constant measured on the Earth. In particular, this parameterization allows for $G_{\rm eff} \ne G$ at present time, $a=1$. This is indeed possible in theories with a screening mechanism~\cite{Vainshtein:1972sx,Damour:1994zq,Khoury:2003aq,Hinterbichler:2010es,Joyce:2014kja}, allowing for gravity to be different cosmologically, while being indistinguishable from GR in the Solar System. As all screening mechanisms are intrinsically nonlinear, they cannot be described using tools of linear perturbation theory employed in MGCAMB. Hence, while testing gravity on linear scales, we effectively assume the existence of a screening mechanism.

{\tt MGCAMB} also allows the user to work with functions $Q(a,k)$ and $R(a,k)$, defined as~\cite{Bean:2010zq}
\begin{align} 
&k^2 \Phi =-4 \pi G \ Q(a,k) \ a^2 \rho \Delta \label{eq:Q} \\
&k^2(\Psi- R(a,k) \Phi) =-12 \pi G \ Q(a,k) \ a^2 (\rho+P) \sigma 
\label{eq:R}
\end{align}

An entirely different parameterization of modified growth, also available in {\tt MGCAMB}, is based on Linder's parameter $\gamma_L$ \cite{Linder:2005in}, defined via $f \equiv \Omega_m(a)^{\gamma_L}$, where $f$ is the growth rate and $\Omega_m(a)$ is the background matter density fraction.

For each of the above choices of phenomenological functions, the user can choose a parameterization from a set of built-in options or add their own. The built-in {\tt MGCAMB} parameterizations can be broadly divided into two categories -- those based on the expressions for $\mu$ and $\gamma$ obtained from scalar-tensor theories in the quasi-static approximation, and the {\it ad hoc} parameterizations introduced in the literature. The latter includes the Planck~\cite{Ade:2015rim} and DES~\cite{DES:2018ufa} parameterizations of $\mu$ and $\Sigma$, while in the former category there are the Bertschinger-Zukin parameterization~\cite{Bertschinger:2008zb}, which applies to most scalar-tensor theories, the generic~\cite{Zhao:2008bn} and the Hu-Sawicki~\cite{Hu:2007nk} $f(R)$, the symmetron~\cite{Hinterbichler:2010es} and the dilaton~\cite{Damour:1990tw,Damour:1994zq,Brax:2012gr}. We note that, in all the scalar-tensor-theory-based parameterizations, it is assumed that baryons and CDM are universally coupled to the scalar field. One of the added features in the 2023 version is the option to constrain scalar field models with coupling only to CDM.

The background evolution in the 2019 version of {\tt MGCAMB} is set by specifying the DE equation of state, $w$, with $w=-1$ corresponding to $\Lambda$, and built-in options for a constant $w$ and the ($w_0,w_a$) parameterization~\cite{Chevallier:2000qy,Linder:2002et}. The 2023 version has an additional option of a parameterization of the DE density fraction $\Omega_X$(a). Note that, in the default {\tt CAMB}, in models with $w \ne -1$, there is a contribution of DE density fluctuations to the Poisson equation calculated either under the assumption of a minimally coupled scalar field~\cite{Weller:2003hw}, {\it i.e.} the quintessence, or the Parameterized Post-Friedmann (PPF) fluid model~\cite{Fang:2008sn}. In {\tt MGCAMB}, $w$ is an effective quantity that need not be associated with a fluid, hence, the DE perturbations were not included. This, however, led to a small discrepancy between the output of {\tt CAMB} and that of {\tt MGCAMB} with $\mu=\gamma=1$ for $w\ne -1$ background models. To give the user the option to eliminate this discrepancy, we added the DE perturbation option to the 2023 version of {\tt MGCAMB}.

\subsection{MGCAMB 2023} 
\label{sec2B}

The new features of this release of {\tt MGCAMB} are:
\begin{itemize}
\item added compatibility with Cobaya, and an implementation in the latest CosmoMC; 
\item added Python wrapper to run {\tt MGCAMB} using the Python interface; 
\item an option to constrain models of a scalar field coupled only to CDM in the QSA limit;
\item a direct implementation of the $\mu$-$\Sigma$ parameterization in the Einstein-Boltzmann solver, eliminating the need to convert to $\mu$-$\gamma$ using Eq.~(\ref{eq:mu-gamma-Sigma});
\item added background model based on parameterization of the DE density (as opposed to $w$), denoted as $\Omega_X$;
\item a non-parametric parameterization of $\mu$, $\Sigma$ and $\Omega_X$, based on a cubic-spline interpolation over a set of discrete nodes in $a$. This allows a joint reconstruction of $\mu$, $\Sigma$ and $\Omega_X$~\cite{Pogosian:2021mcs,Raveri:2021dbu};
\item the option of including DE perturbations to restore the consistency with {\tt CAMB} when working with $w \ne -1$ background models.
\end{itemize}
In what follows, we describe these new features in more detail.

\subsubsection{MGCAMB with Cobaya and CosmoMC}

In order to use Cobaya with {\tt MGCAMB}, one needs to install Cobaya first from the Cobaya website~\cite{cobaya}. Generally, running Cobaya with {\tt MGCAMB} is the same as running it with {\tt CAMB}. We have created an input {\tt YAML} file, available for download at \url{https://github.com/sfu-cosmo/MGCobaya}, that includes both the basic cosmological model parameters and {\tt MGCAMB}-specific new parameters along with complete instructions. Users are referred to the provided template file {\tt temp.yaml} and can modify it according to which MG model they want to work with.

Additionally, a new version of MGCosmoMC, which is a modified version of the latest release of CosmoMC with {\tt MGCAMB} implemented in it, is publicly available at: \url{https://github.com/sfu-cosmo/MGCosmoMC}.

\subsubsection{Scalar field coupled only to CDM} 
\label{sec2B1}

{\tt MGCAMB} evolves the full set of Einstein-Boltzmann equations parameterized via the functions $\mu$-$\gamma$. Built-in expressions for the latter are typically based on their QSA form in scalar-tensor theories and are derived in the Jordan frame, in which baryons and CDM follow the geodesics and obey the standard conservation equations, while the scalar field is coupled to the metric, thus modifying Einstein equations. The Brans-Dicke type theories, such as $f(R)$, can also be formulated in the Einstein frame, conformally related to the Jordan frame, in which Einstein equations are not modified, but all the matter is non-minimally coupled to the scalar field. As all of our observational tools and units are based on the Standard Model physics, theoretical predictions must be made in the baryon frame, which is the Jordan frame in this case.

In addition to the universally coupled case, it is interesting to study models in which the scalar field only couples to CDM~\cite{Amendola:2016saw,Barros:2018efl}. In this case, the baryon frame is the Einstein frame, i.e. Einstein's equations are not modified. Instead, the CDM conservation equations are modified by the coupling to the scalar field. While the cosmological phenomenology of the universal and CDM-only coupled cases is very similar, the technical implementations of the two in {\tt MGCAMB} are different. In the latter case, $\mu=\gamma=\Sigma=1$, but the CDM continuity and Euler equations acquire  new terms. In the Newtonian gauge, in Fourier space, they are given by (see Appendix~\ref{appendixA} for more details)
\ba
\dot{\delta}_c+\theta_c -3 \dot{\Phi}- \dot{(\beta \delta \phi)}  = 0
\label{eq:contic1} \\
\dot{\theta}_c+\left[{\cal H}+\beta \dot{\phi}^{(0)}\right] \theta_c -k^{2} \Psi= \beta k^{2} \delta \phi \ ,
\label{eq:Eulerc1}
\ea
where $\delta_c$ is the density contrast, $\theta_c$ is the velocity divergence, $\dot{ }=\partial / \partial \tau$, ${\cal H} = \dot{a}/a$, ${\phi}^{(0)}$ is the background scalar field, $\delta \phi$ is the perturbation, $m(\phi)$ and $\beta(\phi)$ are the mass and the coupling functions defined in Appendix~\ref{appendixA}. When applying the QSA, we assume that all time-derivatives of the scalar field can be neglected, giving
\ba
\dot{\delta}_c+\theta_c = 0
\label{eq:contic2} \\
\dot{\theta}_c+{\cal H} \theta_c -k^{2} \Psi= \beta k^{2} \delta \phi \ ,
\label{eq:Eulerc2}
\ea
with $\delta \phi$ algebraically related to the density contrast:
\be
\delta \phi=-\frac{\beta \rho_c \delta_c}{k^{2}/a^{2}+m^2} \ ,
\label{eq:deltaphi1}
\ee
thus eliminating the scalar field entirely from all equations. We note that this is a strong version of the QSA which is not applicable to theories in which kinetic energy of the scalar field is a non-negligible fraction of the total energy~\cite{Baldi:2012ua}. However, this is a good approximation for theories like chameleon, symmetron and dilaton, in which the scalar field remains at the minimum of the slowly evolving effective potential. In theories where this is not the case, one should add the scalar field explicitly to the code, which is not done in {\tt MGCAMB}.

To implement in {\tt MGCAMB}, it is necessary to convert the CDM Euler equation to synchronous gauge using~\cite{Ma:1995ey}
\ba
\delta^{\rm (syn)}=\delta^{\rm (con)}-\alpha \frac{\dot{\rho}}{\rho} \label{eq:transf_delta}\\
\theta^{\rm (syn)}=\theta^{\rm (con)}-\alpha k^{2} \label{eq:transf_theta} \ , 
\ea
giving
\ba
\label{eq:continuity-sync}
\dot{\delta}_c+\theta_c + {1\over 2}\dot{h}= 0 \\
\dot{\theta}_{c}+{\cal H}{\theta}_{c}=-k^{2} \tilde{\beta}^{2} \frac{\rho_{c}\left(\delta_{c}-3 \alpha {\cal H} \right)}{k^{2}/a^{2}+m^{2}}
\label{eq:Eulerc3_QSA}
\ea
where all the quantities are now in synchronous gauge, $\tilde \beta(\phi) =  \beta(\phi)/\sqrt{8\pi G}$, $\alpha=(\dot{h}+6 \dot{\eta}) / (2 k^{2})$, and $h$ and $\eta$ are the synchronous gauge potentials~\cite{Ma:1995ey}. 

The 2023 {\tt MGCAMB} has built-in parameterizations for CDM-coupled scalar field models based on the forms of $m(a)$ and $\tilde \beta(a)$ in the symmetron and dilaton models as described in~\cite{Hojjati:2015ojt}. In~\cite{Mirpoorian:2023utj}, we tested the validity of the QSA for these models by comparing to the exact solutions with an explicitly present scalar field, and found that the QSA works very well for a broad range of parameters. Other forms of $m$ and $\tilde \beta$ are straightforward to add.

\subsubsection{Direct $\mu-\Sigma$ parametrization} 
\label{sec2B2}

As mentioned earlier, the original version of {\tt MGCAMB} was based on the set of linearly perturbed Einstein equations parameterized via the functions $\mu(a,k)$ and $\gamma(a,k)$. Depending on the data sets that one considers, it can be preferable to work with the combination $\mu(a,k)$ - $\Sigma(a,k)$. In the original version, this choice would be first transformed into the corresponding  $\mu(a,k)$ - $\gamma(a,k)$, with {\tt MGCAMB} determining $\gamma$ from $\gamma = 2\Sigma/\mu -1$, along with the derivatives. As discussed earlier, the latter is an approximated relation valid as long as the modifications occur well-after the onset of matter domination and anisotropic stresses of relativistic species can be neglected. It can be generalized to properly take into account these effects; nevertheless, also in view of numerical accuracy, we opted for adding a direct implementation of $\mu$ and $\Sigma$, with the latter defined via Eq~(\ref{def:Sigma}), in the equations for perturbations. When working with $\mu-\Sigma$ models, the user can opt for this direct implementation by setting {\tt MG\_flag=5}. Using other values for {\tt MG\_flag} would revert to the old way based on converting to $\gamma$.

The modifications to equations in the $\mu-\Sigma$ case are similar to those for $\mu-\gamma$ described in~\cite{Zucca:2019xhg}. Here we point out the main differences. The modified Poisson equations with the $\mu-\Sigma$ parameterization are defined in the Newtonian gauge via Eqs~(\ref{eq:Poisson1}) and (\ref{def:Sigma}). To convert to the synchronous gauge, and to find the variable $z$ used in {\tt CAMB},
\begin{equation}
 z = k \alpha - 3 \frac{\dot{\eta}}{k} \ ,
 	\label{eq:z}
\end{equation}
we start with the transformations given by Eqs.~\eqref{eq:gauge1} and \eqref{eq:gauge2} to find $\alpha$ and $\dot{\alpha}$ as
\begin{align}
&\dot{\alpha}=-\eta-\frac{a^{2}}{2 k^{2}} \Sigma \Big[2 \rho \Delta+3 \rho \big(1+w\big) \sigma \Big]  \label{eq:alpha} \\
&\alpha=\frac{1}{{\cal H}} \Big\{\eta+\frac{a^{2}}{2 k^{2}} \Big[ \big(2\Sigma-\mu\big) \rho \Delta +\big(\Sigma-\mu\big) 3 \rho\big(1+w\big)\sigma \Big] \Big\}.
\label{eq:alpha,alphadot}
\end{align}
Then, following the same steps that were used in the $\mu-\gamma$ case, as described in~\cite{Zucca:2019xhg}, we obtain 
\begin{align}\label{eq:etadot}
\dot{\eta}&=\frac{1}{2} \frac{a^{2}}{k^{2}+\frac{3}{2} a^{2}(2\Sigma-\mu) \rho(1+w)}\nonumber \\
&\Big\{(2\Sigma-\mu)k  \rho q\Big[1+\frac{3({{\cal H}}^2-\dot{{\cal H}})}{k^{2}}\Big] \nonumber \\
&+\rho \Delta \big[2{\cal H}(\Sigma-\mu)-(2\dot{\Sigma}-\dot{\mu})\big] \nonumber\\ 
&+ k^{2} \alpha\Big[(2\Sigma-\mu)  \rho(1+w)-\frac{2}{a^2}(H^2-\dot{{\cal H}})\Big] \nonumber \\
&- 2(\Sigma-\mu) \rho \dot{\Pi} + 2 \rho \Pi \big[(\Sigma-\mu)3{\cal H}(1+w) - (\dot{\Sigma}-\dot{\mu})\big]\Big\},
\end{align}
where $\Pi=\frac{3}{2}(1+w) \sigma$, and $(1+w) \theta=kq$, allowing us to determine $z$.

We have tested that, for late-time modifications (which includes all models currently implemented in {\tt MGCAMB}), the results are equivalent to those based on the conversion to $\gamma$. Still, the added option gives the users the possibility to work with new models that may involve early-time modifications of $\mu$ and $\Sigma$.

\subsubsection{Effective dark energy density fraction $\Omega_X$}

The dynamics of DE in {\tt CAMB} is set by specifying the equation of state parameter $w$. In this version of {\tt MGCAMB}, we added the option of specifying the DE energy density instead. Namely, we introduce a function $\Omega_{\mathrm{X}}(a)$, defined 
via the Friedmann equation, 
\be
\frac{H^2}{H_0^2}=\Omega_{r}a^{-4}+\Omega_{m}a^{-3} + \Omega_{\mathrm{X}}(a) \ ,
\label{eq:omega_X}
\ee
where $H=a^{-1}da/dt$ is the Hubble parameter (defined in terms of the physical time $t$), $H_0$ is its present value, $\Omega_{r}$ and $\Omega_{m}$ are the fractional energy densities of radiation and matter, respectively, and $\Omega_X = \Omega_{\mathrm{DE}}X(a)$ with $X(a=1)=1$ and $\Omega_r+\Omega_m+\Omega_{\mathrm{DE}}=1$. Thus, $\Omega_{\mathrm{X}}(a)$, describes the collective contribution of any terms other than the radiation and matter densities, including terms due to modifications of gravity that may alter the Friedmann equation. In MG theories, the energy density of the effective DE fluid, defined as above, need not be positive and can cross zero, making its equation of state singular. Hence, when deriving constraints on MG, it is reasonable to avoid introducing $w$ and work with the effective DE density instead.
 
The effective DE pressure $p_{\mathrm{DE}}^{\mathrm{eff}}$, used in the equations in {\tt CAMB}, can be obtained from $X(a)$ via~\cite{Wang:2018fng}
\be
Y=-X-\frac{1}{3} \frac{dX}{da} a 
\label{eq:Y}
\ee
with $Y(a) = p_{\mathrm{DE}}^{\mathrm{eff}}(a) / \rho_{\mathrm{DE}}^{\mathrm{eff}}(a=1)$.

In the current version of {\tt MGCAMB}, $\Omega_X(a)$ is implemented as a cubic spline over a discrete set of nodes, as detailed in the next section. Other parameterizations can be added following the general scheme for adding new models to {\tt MGCAMB}. 

To work with $\Omega_X$, one needs to choose {\tt DE\_model = 3} in the {\tt params\_MG.ini} file. This option can be used independently from the choice of parameterizations of $\mu$ and $\Sigma$. For example, to use the ``DES'' parameterization along with $\Omega_X$, one needs to set {\tt MG\_flag = 1}, {\tt pure\_MG\_flag = 2}, {\tt musigma\_par = 1} and {\tt DE\_model = 3}.

\subsubsection{The cubic-spline parameterization and reconstructions} 
\label{sec2B3}

To allow for a non-parametric reconstruction of the functions $\mu$, $\Sigma$ and $\Omega_X$, in this release, we provide an implementation of a pixelization which can be easily modified or extended by the user for their own purpose.

We restrict to the time-dependent case, and parameterize the functions $\mu$, $\Sigma$ and $\Omega_X$ by fitting nodes placed at certain chosen redshifts, and the values of these functions at intermediate redshifts are determined by a cubic spline interpolation. In the code, each function is assigned with $11$ fitting nodes, with the first ten uniformly distributed in the redshift range of $z\in$ [0,3], and the last one set at $z=4$. This is because most observations can only provide tomographic measurements at $z<3$, thus variations of these functions beyond $z=3$, if any, are difficult to probe. From $z=4$ to $z=1000$, we require all three functions to smoothly transit from the fitted value at $z=4$ to the $\Lambda$CDM value, which are unity for $\mu$, $\Sigma$, and $\Omega_{\mathrm{DE}}$ for $\Omega_X$. This is hardcoded using $9$ additional nodes uniform in the scale factor $a$ whose values are determined by a ${\tt tanh}$ function. In sum, there are $11\times3-1=32$ free parameters\footnote{$X(a=1)=1$ by definition.} to be determined, which is quite challenging given the strong degeneracies among these parameters. The way out is to apply the correlated priors~\cite{Crittenden:2005wj,Crittenden:2011aa}, which can be calculated in theory~\cite{Raveri:2021dbu,Pogosian:2021mcs}, to remove the flat directions of the likelihood surface. The covariance matrices for a few correlated priors are available in the MGCosmoMC package: \url{https://github.com/sfu-cosmo/MGCosmoMC}, under {\tt data/corr\_prior}. To make the priors work with Cobaya, one would need to construct an external likelihood class following the general way instructed on the Cobaya website~\cite{cobaya}. The priors are simply implemented as a new contribution to the total $\chi^2$ via:
\be
\chi^2= \left(\mathbf{f}-\mathbf{f}_{\text{fid}}\right) \mathcal{C}^{-1}\left(\mathbf{f}-\mathbf{f}_{\mathrm{fid}}\right)^{T}
\label{eq:chi-corr}
\ee
where $\mathbf{f} \equiv \left\{\Omega_{X i}, \mu_{i}, \Sigma_{i}\right\}$ describes the discrete nodes for the functions, with the fiducial value $\mathbf{f}_{\text{fid}}$ determined by the so-called running average method~\cite{Silvestri:2013ne} to avoid the statistically biased result, instead of the mean value obtained from the covariance matrices for the functions, which is denoted as $\mathcal{C}$ in the expression.

Note that these three functions do not have to be in the same parametric form. For example, when $\mu$ and $\Sigma$ are parameterised using the aforementioned fitting nodes with the cubic spline, dark energy can take the ($w_0, w_a$) parametrization, in which case one needs to set {\tt MG\_flag = 6} and {\tt DE\_model = 2} in the {\tt params\_MG.ini} file. Likewise, when $\Omega_X$ is a free function with fitting nodes, $\mu$ and $\Sigma$ can take a simpler form as already mentioned in the previous subsection.

\subsubsection{DE perturbations} 
\label{sec2B4}

A dynamical DE, {\it i.e.} any form of DE other than $\Lambda$, necessarily implies inhomogeneities in the DE fluid~\cite{Caldwell:1997ii}. In {\tt CAMB}, in models with $w \ne -1$, the DE stress-energy fluctuations are computed either based on a quintessence scalar field~\cite{Weller:2003hw} or the PPF fluid model~\cite{Fang:2008sn}. In previous versions of {\tt MGCAMB}, the DE contribution to the stress-energy perturbations was not included in the perturbed Einstein equations. Instead, it was assumed that their contribution would be absorbed into the phenomenological functions $\mu$ and $\gamma$. This, however, caused a small but noticeable difference between the best fit parameters obtained using {\tt CAMB} compared to the $\mu=\gamma=1$ limit of {\tt MGCAMB} for $w \ne -1$ background cosmologies. In the current version, we have added the option {\tt MGDE\_pert} in the {\tt params\_MG.ini} to include the DE perturbations, calculated using the quintessence or the PPF model, in the equations of {\tt MGCAMB}, in the same way as they appear in {\tt CAMB}. Whether the DE perturbations should be included when running {\tt MGCAMB} depends on the context. 

In MG theories, the DE equation of state $w$ is not necessarily representative of a scalar field, or a conserved fluid assumed by the PPF model. Rather, it is an effective quantity representing the overall modification of the Friedman equation due to changes to the Einstein equation as well as the impact of inhomogeneities in the extra degree of freedom. Hence, using the quintessence or the PPF model for DE perturbation when performing model-agnostic tests of MG is, strictly speaking, theoretically inconsistent. On the other hand, if the user's priority is to be able to recover the default-{\tt CAMB}-based results in the $\mu=\gamma=\Sigma=1$ limit when running $w\ne -1$ models, they should include the DE perturbations.

In addition to the DE perturbations arising from DE dynamics, in scalar-tensor theories, in the Einstein frame, there is a contribution to the energy density perturbations in the Poisson equation due to the non-minimal coupling of the scalar field to matter. Since the baryon frame in the CDM-only coupled case is the Einstein frame, this term appears on the right hand side of Eq.~(\ref{eq:psi}). We note that it is generally very small on sub-horizon scales and could be safely neglected for models for which the QSA holds well. Nevertheless, we keep it for completeness.

\section{Testing gravity with the new MGCAMB} 
\label{sec3}

In what follows, we demonstrate the use of the new {\tt MGCAMB} for deriving constraints on parameters of two new models that were not present in previous versions. The first is the CDM-only coupled symmetron, which we compare to the universally (all matter) coupled case. The second is a joint reconstruction of $\Omega_X$, $\mu$ and $\Sigma$ as functions of redshift using the cubic spline model with and without the Horndeski prior. 

\subsection{Constraints on the universally and the CDM-only coupled symmetron}

\begin{figure}
\centering
\includegraphics[width = .5\textwidth]{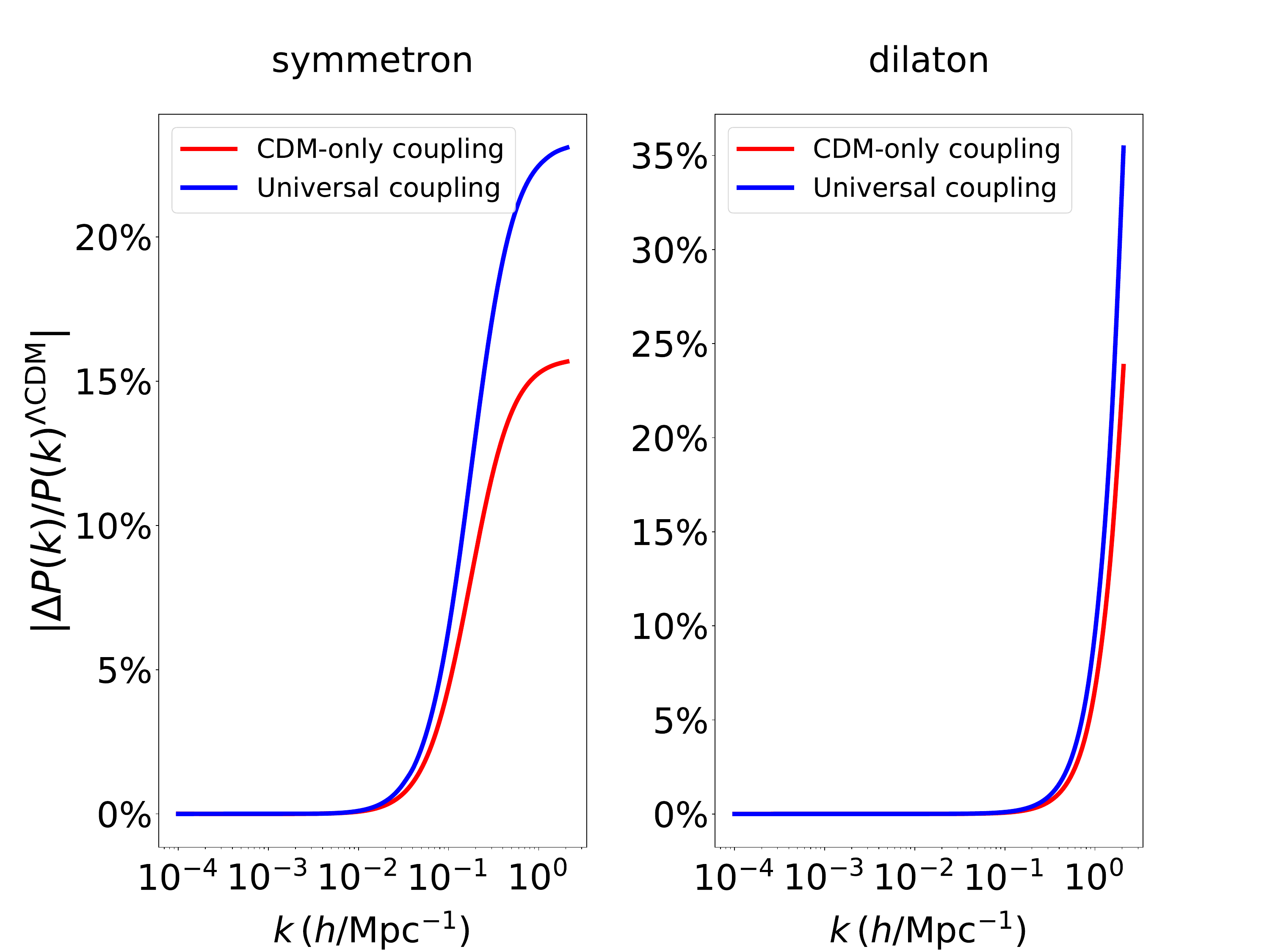}
\caption{\label{fig:pk2}The relative difference in the matter power spectrum for two scalar field models with universal and CDM-only coupling compared to $\Lambda$CDM model. For the symmetron model, the parameter values are $\beta_*=1.0$, $\xi_* = 10^{-3}$ ($\lambda_c=4.29$~$\mathrm{Mpc}$), $a_*=0.5$; while the parameter values for dilaton model are $\beta_{0}=1.0$, $\xi_{0}=10^{-4}$ ($\lambda_c=0.43$~$\mathrm{Mpc}$), $a_{\rm trans}=0.001$, where $a_{\rm trans}$ sets the transition time from GR to MG regime. In both panels, the blue line corresponds to the CDM-only coupled scalar field, while the red line corresponds to the universally coupled scalar field.}
\end{figure}

The built-in parameterizations in {\tt MGCAMB} include the symmetron and dilaton models described in terms the QSA forms of the mass function $m(a)$ and the coupling $\tilde \beta(a)$. As described in Sec.~\ref{sec2B1}, in addition to the previously implemented case of the scalar field universally coupled to all matter, the current version also includes the option for the scalar field coupled only to CDM. Fig. \ref{fig:pk2} compares the matter power spectra of two scalar field models, symmetron and dilaton, with CDM-only {\it vs} universal coupling. In both cases and for both models there is an enhancement due to the fifth force. However, this enhancement is smaller in the CDM-only coupled case, since the fifth force affects only a fraction of total matter.

In what follows, we compare the constraints on the symmetron model for the two cases with the universal and the CDM-only coupling. To work with this model, we set {\tt MG\_flag=3, QSA\_flag = 2} for the universal coupling and {\tt MG\_flag=4, QSA\_flag = 2, CDM\_flag = 1} for the CDM-only coupling in the {\tt params\_CMB\_MG.ini} file for the CosmoMC runs, or the input {\tt YAML} file if using Cobaya.

In the symmetron model, under the QSA, the functions $\tilde \beta(a)$ and $m(a)$ are given by~\cite{Brax:2012gr,Hojjati:2015ojt}
\ba
\tilde{\beta}(a)=\beta_{\star} \sqrt{1-\left(\frac{a_{\star}}{a}\right)^{3}}  \\
m(a) = \frac{H_0}{c}\frac{1}{\xi_{\star}} \sqrt{1-\left(\frac{a_{\star}}{a}\right)^{3}} \ ,
\ea
for $a > a_{\star}$, where $a_*$ is the scale factor at which the symmetry breaking takes place. Prior to the symmetry breaking, the minimum of the effective scalar field potential is at $\phi=0$, implying $\tilde{\beta}(a)=0$. Hence, under the QSA, the symmetron model reduces to $\Lambda$CDM, and we evolve the $\Lambda$CDM equations for $a < a_{\star}$. In~\cite{Mirpoorian:2023utj}, it was found that these QSA-based expressions work well for a broad range of parameters. 

Current data is unable to constrain all three symmetron parameters ($\xi_*$, $\beta_*$ and $a_*$) simultaneously. For our demonstration, we set  $\beta_*=1$ and $a_*=0.5$, and constrain the remaining parameter $\xi_*$ that sets the Compton wavelength of the fifth force mediated by the scalar field,
\begin{equation}
\lambda_c = \frac{c}{H_0} \xi_* \ .
\label{eq:lambda_c}
\end{equation}
We present our results in terms of $\lambda_c$.

We use the new version of {\tt MGCosmoMC} to compare constraints on $\lambda_c$ in the symmetron model with universal and CDM-only coupling. Along with $\xi_*$, we vary the main cosmological parameters: $\Omega_bh^2$, $\Omega_ch^2$, $\theta$, $\tau$, $n_s$ and $\ln[10^{10}A_s]$. Our dataset includes the Planck 2018 CMB temperature, polarization and lensing~\cite{Aghanim:2019ame}, joint measurements of baryon acoustic oscillations (BAO) and redshift-space distortions (RSD) from BOSS DR12~\cite{BOSS:2016wmc}, the SDSS DR7 MGS data~\cite{Ross:2014qpa}, the BAO measurement from 6dF~\cite{Beutler_2011}, and the Dark Energy Survey (DES) Year 1 galaxy-galaxy-lensing correlation data~\cite{Abbott:2017wau} with the standard cut of nonlinear scales (see \cite{Zucca:2019xhg} for more details on the implementation of the cut). We use a logarithmic prior on $\xi_*$ covering seven orders of magnitude, $\xi_* \in [10^{-6}, 1]$.

\begin{table*}[t]
 \centering
 \begin{tabular}{lcccc}
  \hline
  coupling & Universal coupling & CDM-only coupling & Universal coupling+$A_L$ & CDM-only coupling+$A_L$\\
  \hline
  $\xi_*$ & 0.0011 & 0.0016 &  0.0013 & 0.0022\\
  $\lambda_c$(Mpc) & 4.743 & 7.051 & 5.802 & 9.539\\
  \hline
 \end{tabular}
\caption{The 2$\sigma$ upper bounds on $\xi_*$ and $\lambda_c$ for the two models with fixed and varying $A_L$, respectively, as defined in the text.}
\label{tab:bound}
\end{table*}

\begin{figure}
\centering
\includegraphics[width = .45\textwidth]{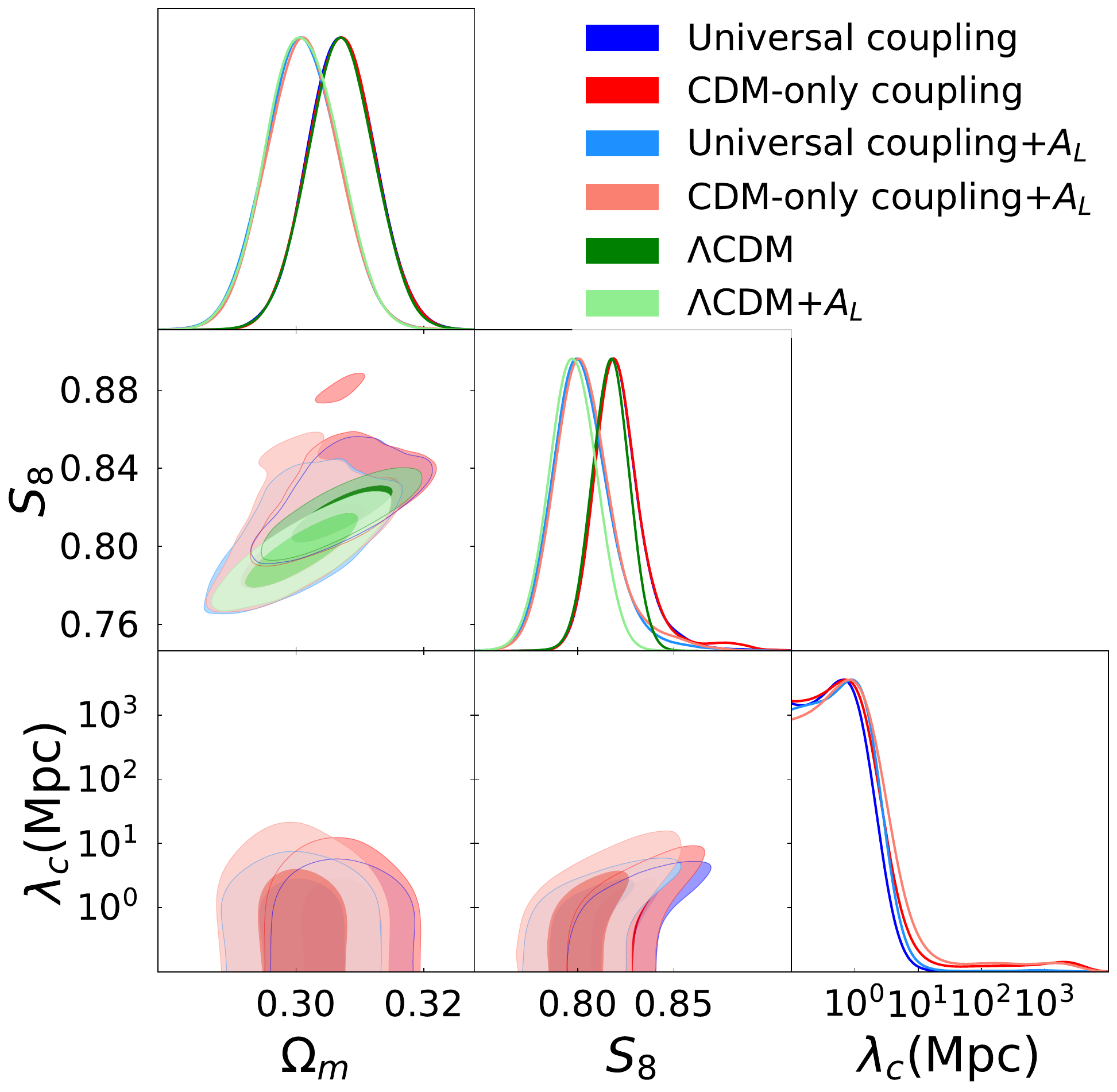}
\caption{\label{fig:lambda_c}
The 68\% and 95\% marginalized confidence level contours for $\Omega_m$, $S_8$ and $\lambda_c$ in the universally coupled (blue) and the CDM-only coupled symmetron (red), with fixed and varying $A_L$, respectively. The other symmetron model parameters are fixed at $\beta_*=1$ and $a_*=0.5$. The $\Lambda$CDM (green) contours are shown for comparison.}
\end{figure}

Fig.~\ref{fig:lambda_c} shows the marginalized joint constraints on $\lambda_c$ for the universal and the CDM-only coupling cases, along with the derived parameters $S_8$ and $\Omega_m$. As expected, $\lambda_c$ is constrained more stringently in the universally coupled case compared to the CDM-only coupled case, whether the CMB lensing parameter, $A_L$, is fixed or set free to vary, since all of the matter is affected by the fifth force in the universally coupled case. The quantitative 68\% and 95\% confidence level constraints on $\lambda_c$ cannot be readily obtained from {\tt getdist} due to the fact that we use a logarithmic prior and the parameter is unbounded from below. Instead, in Table~\ref{tab:bound}, we provide the ``2$\sigma$'' upper bounds on $\xi_*$ and $\lambda_c$ defined as the value of the parameter at which the marginalized probability is equal to $1/e^2$ of the peak value.  For a Gaussian distribution, this would set the $95$\% confidence level, or the $2\sigma$ bound.

We note that there is a degeneracy between $S_8$ and $\lambda_c$ at larger values of $S_8$, which is plausible since the fifth force tends to increase the clustering amplitude of matter. Note that the value of $\Omega_m$ stays the same, {\it i.e.} the increase in clustering can be achieved without increasing the matter fraction. In addition, the mean values of $\Omega_m$ and $S_8$ are lower when $A_L$ is varying, since CMB temperature data tends to elevate $\Omega_m$ and $S_8$ due to the preference for more CMB lensing effects when $A_L$ is fixed. Instead, when $A_L$ is varied, one finds a preference for $A_L>1$.

\subsection{Reconstructing gravity with and without a Horndeski prior}

As a second worked out example, we perform a combined non-parametric reconstruction of $\Omega_X$, $\mu$ and $\Sigma$ from current cosmological data.  We set {\tt \{MG\_flag=6, DE\_flag=3\}} in the input {\tt params\_CMB\_MG.ini} file, which corresponds to modeling all three functions through the cubic spline over $11$ fitting nodes in redshift, as described in Section~\ref{sec2B3}. We  fit the resulting $32$ free parameters of the theory, along with the standard cosmological parameters, to data, considering both the case with and without the Horndeski correlated prior (as already discussed, the Horndeski prior can be added by including {\tt cor\_prior\_Hor\_ox.ini} and setting {\tt use\_SMPrior = T} in the input {\tt .ini} file). 

\begin{figure}
\centering
\includegraphics[width = .46\textwidth]{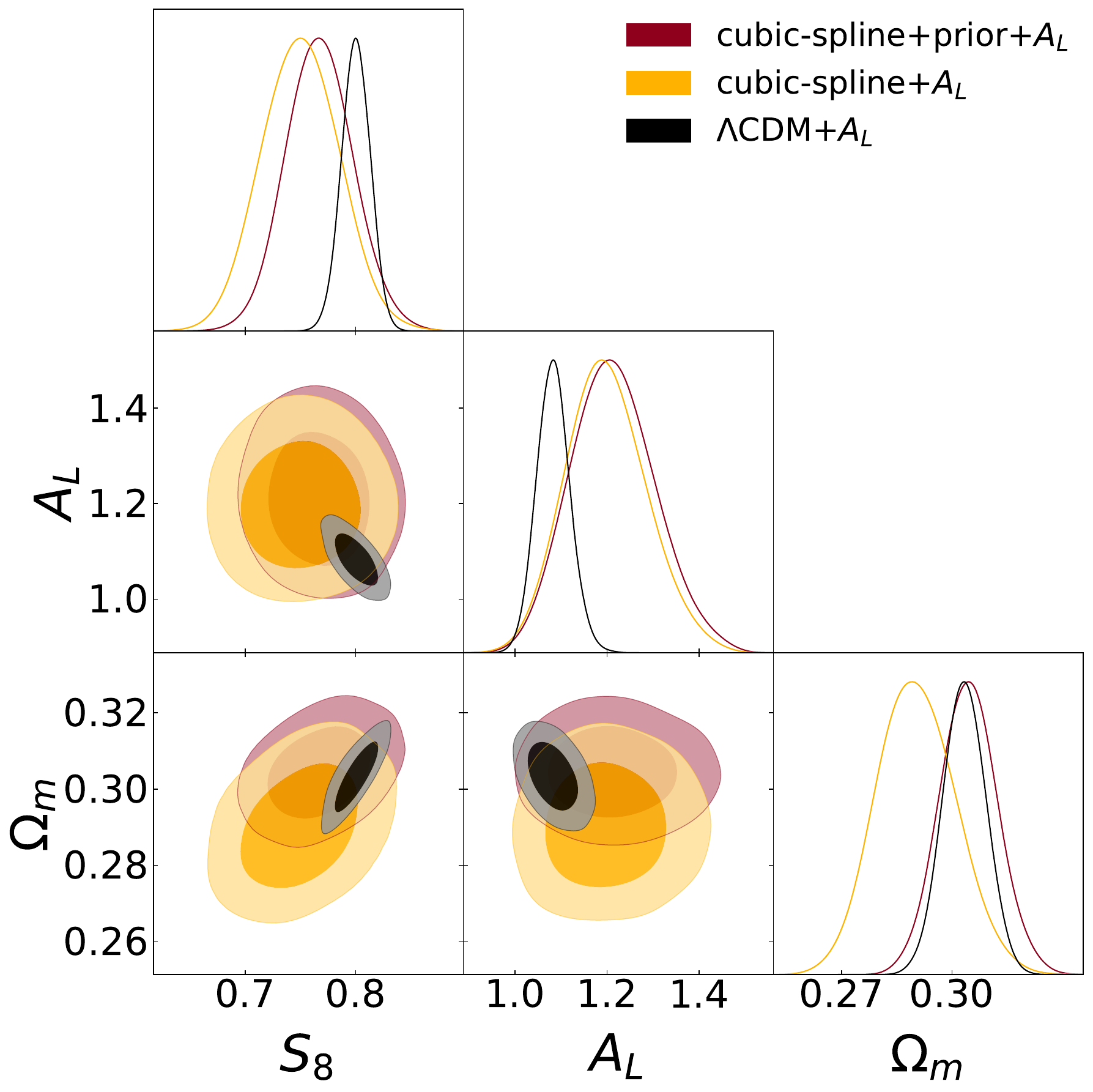}
\caption{\label{fig:recon_tri}The 68\% and 95\% marginalized confidence level contours for $S_8$, $A_L$ and $\Omega_m$, in the cases of joint reconstruction of $\mu$, $\Sigma$ and $\Omega_X$ using the same datasets with varying $A_L$, with and without the Horndeski prior, respectively, in comparison with $\Lambda$CDM model.}
\end{figure}

\begin{figure}
\centering
\includegraphics[width = .46\textwidth]{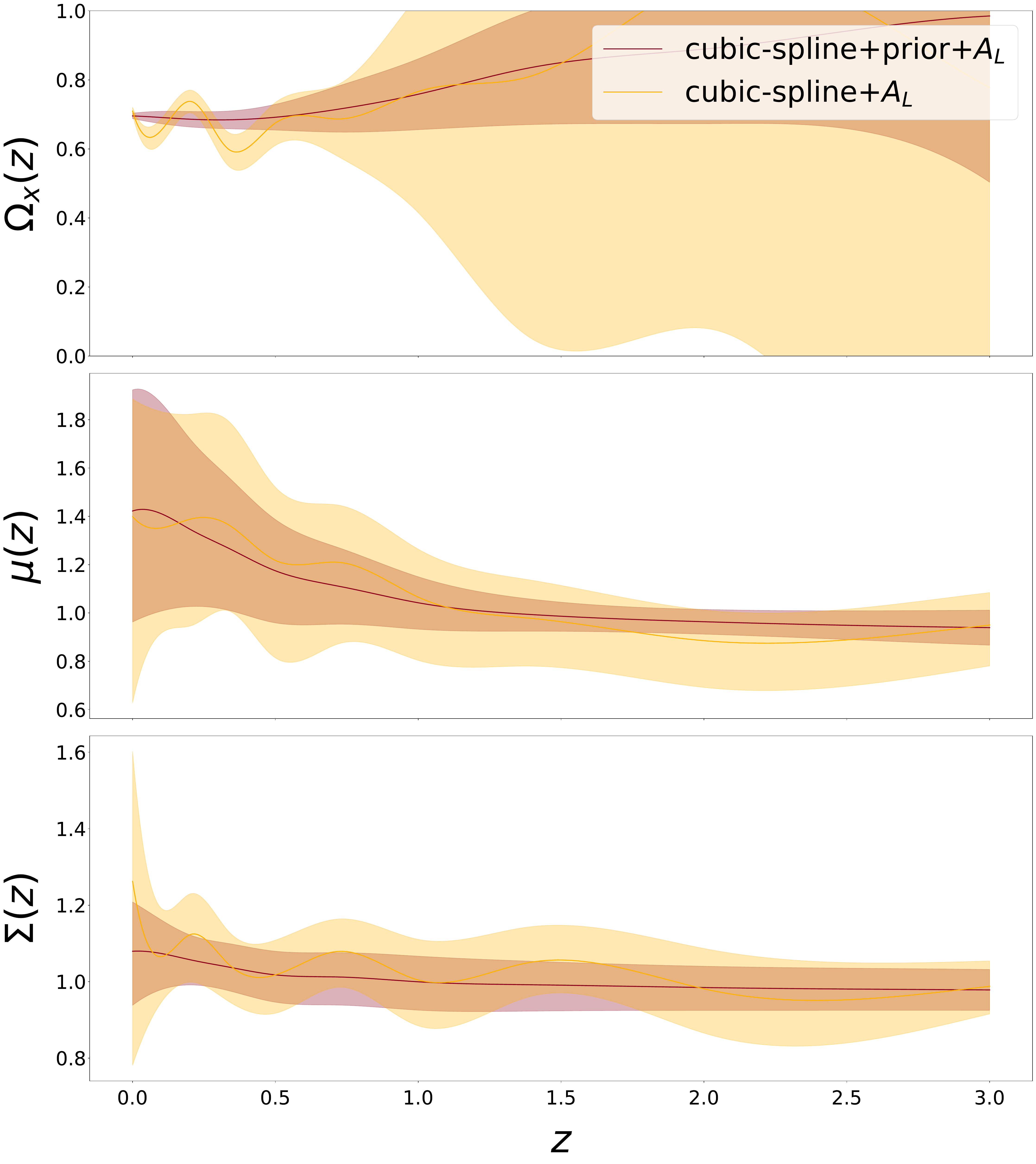}
\caption{\label{fig:recon_mg}The evolution of $\mu$, $\Sigma$ and $\Omega_X$ derived from the reconstruction by cubic-spline model, using all data with varying $A_L$, with and without the Horndeski prior. The bands in all subplots correspond to the 68$\%$ confidence level regions obtained from the marginalized posterior distributions for each node, and the solid lines inside the bands correspond to the mean values of parameter functions.}
\end{figure}

Our dataset is comprised of the Planck 2018 CMB temperature, polarization and lensing spectra~\cite{Aghanim:2019ame}, the full shape consensus results of joint measurements of BAO and RSD from BOSS DR12~\cite{BOSS:2016wmc} complemented by portion of the eBOSS DR16 data release~\cite{Zhao:2020tis,Hou:2020rse,Bautista:2020ahg,deMattia:2020fkb,Neveux:2020voa,Wang:2020tje,duMasdesBourboux:2020pck} not included in DR12, the BAO measurements from MGS and 6dF, the Pantheon sample of uncalibrated supernovae~\cite{Pan-STARRS1:2017jku}, along with the DES Y1 data with the standard cut of nonlinear scales. The results are shown in  Fig.~\ref{fig:recon_tri} and Fig.~\ref{fig:recon_mg}. They reproduce the more general results obtained in~\cite{Pogosian:2021mcs,Raveri:2021dbu}, where the same set of data were used. We refer the reader to the latter work for an extensive discussion of the findings. Here we shall simply comment on the main points: the reconstructed functions are consistent with their $\Lambda$CDM predictions within $2-3\,\sigma$; the role of the prior, in preventing overfitting of the data, is clearly visible; all three functions show some mild deviations from their $\Lambda$CDM values, hinting at the features that would be needed for late time dark energy in order to ease some of the cosmological tensions~\cite{Abdalla:2022yfr}. In particular, as can be seen in Fig.~\ref{fig:recon_tri}, it is possible for late time modifications to ease the $S_8$ tension if $A_L$,  is let free to vary. This is achieved mostly through the combined behavior of $\Omega_X$ and $\mu$, with an increase in $\Omega_X$ at intermediate redshifts, and  $\mu$ achieving values above unity at low and intermediate redshifts, while $\Sigma$ is close to unity in the redshift range relevant for the CMB lensing kernel.

One expects the values of $\mu$, $\Sigma$ and $\Omega_X$ at adjacent redshifts not to be entirely independent, as these functions are generally smooth and correlated with each other. The Horndeski correlated prior plays an important role in suppressing the likelihood of abrupt unphysical changes in the data-only reconstruction, preventing an overfitting and ensuring that the reconstruction is independent of the binning scheme. We use the prior covariance obtained from the ensemble of cosmological histories within Horndeski theories generated in~\cite{Espejo:2018hxa}. There, the space of Horndeski models was sampled by varying the five free functions of time that appear in the effective field theory (EFT) action for (linear) Horndeski gravity, keeping only the solutions that satisfy basic principles of physical viability ({\it e.g.} no instabilities) and are in broad agreement with the observed cosmic expansion history.  The strength of the resultant correlation depends on the amount of freedom in a given model ({\it e.g.} the entire Horndeski {\it vs} only the Brans-Dicke subset). We have made the covariance matrices for a few correlated priors  available in the {\tt MGCosmoMC} package, but the user could follow the same template to build other correlation matrices, e.g. for other theories of gravity. 
 
 While it is not evident from the results that we are presenting here, in the original reconstruction work~\cite{Raveri:2021dbu} it was shown that current cosmological data can constrain $15$ combined modes of $\Omega_X$, $\mu$ and $\Sigma$. This is already significantly more than the few parameters typically employed in simple parameterizations and highlights the importance of non-parametric methods.  

\section{Summary}
We have presented a new version of {\tt MGCAMB} which is now publicly available at \url{https://github.com/sfu-cosmo/MGCAMB} and comes with tools for using it with {\tt Cobaya}~\cite{Lewis:2002ah,Torrado:2020dgo,cobaya}, as well as an implementation in the latest version of {\tt CosmoMC} ~\cite{Lewis:2002ah,Lewis:2013hha, cosmomc}. This new version includes several added features: new built-in models, {\it e.g.}  a scalar field coupled only to dark matter in the quasi-static approximation; a direct implementation of the $\mu$-$\Sigma$ parameterization in the Einstein-Boltzmann solver, eliminating the need to convert to $\mu$-$\gamma$; the option to include DE perturbations when working with $w\ne -1$ backgrounds; a new binned parameterization allowing for a simultaneous reconstruction of $\mu$, $\Sigma$ and $\Omega_X$ (the fractional dark energy density) as functions of redshift. For each of these we have provided detailed instructions on how to use it. In some cases, we have also provided a worked out example complete with fits to currently available cosmological data. 

This new version brings {\tt MGCAMB} up to speed with the more recent analysis tools developed in preparation for the upcoming cosmological surveys. It makes it also more versatile and complete in terms of theoretical scenarios that can be explored and resolves the small glitch w.r.t. the output of {\tt CAMB} for cosmologies with $\mu=1=\gamma$ but $w\neq-1$. This feature comes with some warnings about when to activate it in order to maximize theoretical consistency. 

Of particular importance in view of upcoming LSS surveys, is the new possibility to run  {\tt MGCAMB} directly with a $\mu-\Sigma$ parameterization of the Einstein equations, without it converting to $\mu-\gamma$. This will improve precision in the analysis for which $\Sigma$ is a more obvious function to fit to data, such as in weak lensing surveys.  

As it was shown in~\cite{Pogosian:2021mcs,Raveri:2021dbu}, cosmological data can constrain much more than the few parameters typically employed in simple parameterizations of dark energy and modified gravity. The constraining power will further increase with upcoming LSS surveys and the new built-in feature of {\tt MGCAMB}, with the simultaneous binning  of $\mu$, $\Sigma$ and $\Omega_X$, will facilitate interesting reconstructions. 

\acknowledgments 
We thank Matteo Martinelli, Marco Raveri, Ruiyang Zhao and Alex Zucca for their valuable help. The work of ZW, HM and LP is supported by the National Sciences and Engineering Research Council (NSERC) of Canada. AS acknowledges support from the NWO and the Dutch Ministry of Education, Culture and Science (OCW). GBZ is supported by the National Key Basic Research and Development Program of China (No. 2018YFA0404503), NSFC (11925303, 11890691), and science research grants from the China Manned Space Project with No. CMS-CSST-2021-B01.

\appendix

\section{Scalar field coupled to CDM}
\label{appendixA}

The action of a theory with a scalar field coupled to CDM can be written as
\begin{align}
S=\int \mathrm{d}^{4} x & \big\{\sqrt{-g}\Big[\frac{M_{\mathrm{Pl}}^{2}}{2} \mathcal{R}-\frac{1}{2} \partial_{\mu} \phi \partial^{\mu} \phi-V(\phi)\Big] \nonumber \\ 
& +\mathcal{L}_{c}\left(\psi_{c}, A^2(\phi) g_{\mu \nu}\right)+\mathcal{L}_{\mathrm{SM}}\left(\psi_{\mathrm{SM}}, g_{\mu \nu}\right)\big\}
  \label{eq:action}
\end{align}
where $\mathcal{L}_{\mathrm{SM}}$ represents the Lagrangian density of the Standard Model of particle physics, which includes baryons and radiations, and $\mathcal{L}_{c}$ represents the Lagrangian density of dark matter. Here $M_{\mathrm{pl}} \equiv(8 \pi G)^{-1/2}$ is the reduced Planck mass, and $\phi$ is the scalar field.
The coupling of the scalar field to dark matter arises due to the conformal factor $A(\phi)$ that alters the gravitational metric felt by the CDM. Varying the action in Eq. \eqref{eq:action} with respect to $\phi$ gives the equation of motion for $\phi$,
\begin{equation}
\square \phi=V_{, \phi}-\frac{A_{,\phi}}{A} T^{c}=V_{, \phi}+\frac{A_{,\phi}}{A} \rho_c \equiv V^{\rm eff}_{, \phi}
\label{eq:phiEoM}
\end{equation}
where $_{, \phi}$ is the derivative with respect to $\phi$, $T^{c} = -\rho_c$ is the trace of the CDM stress-energy $T_{\mu \nu}^{c}$, and we have defined the effective potential $V^{\rm eff}$. 
The Einstein's equation is obtained by varying the action with respect to $g_{\mu \nu}$:
\begin{equation}
G_{\mu \nu} = M_{\mathrm{Pl}}^{-2}\left[T_{\mu \nu}^{\rm SM}+T_{\mu \nu}^{c}+T_{\mu \nu}^{\phi}\right] \ ,
\label{eq:Einstein}
\end{equation}
where the stress-energy tensor of the standard matter is conserved, $\nabla^{\mu}T_{\mu \nu}^{\rm SM}=0$, and so is the sum of the scalar field and the CDM stress-energies: $\nabla^{\mu}[T_{\mu \nu}^{c}+T_{\mu \nu}^{\phi}]=0$, but 
$T_{\mu \nu}^{c}$ and $T_{\mu \nu}^{\phi}$ are not individually conserved. We have
\be
\nabla^{\mu} T_{\mu \nu}^{c} =  - \beta \rho_{c} \partial_\nu \phi \ ,
\label{eq:conservationCDM}
\ee
where $\beta \equiv A_{,\phi}/A$.

Perturbing Eq.~(\ref{eq:conservationCDM}) to first order and transforming to Fourier space, yields the continuity and the Euler equations for CDM given by Eqs.~(\ref{eq:contic1}) and (\ref{eq:Eulerc1}). Also, perturbing Eq.~(\ref{eq:phiEoM}) to first order and taking the QSA, gives
\begin{equation}
\delta \phi=-\frac{\beta \rho_c \delta_c}{k^{2}/a^{2}+m^2}
\label{eq:deltaphi2}
\end{equation}
where $m^2 \equiv V^{\text{eff}}_{,\phi \phi}$ evaluated at the minimum of the effective potential. Namely, when applying the QSA we assume that the scalar field is always at the minimum of $V^{\rm eff}$. This allows us to eliminate the scalar field entirely from all equations.

The form of the Einstein equations is unchanged in this model, hence $\mu=\gamma=\Sigma=1$. There is a contribution from the scalar energy density, $V_{,\phi} \delta \phi$, on the right hand side of the Poisson equation, which is negligible for the class of models for which our QSA is valid. However, we include it in the Poisson equation, after using (\ref{eq:deltaphi2}), as
\be
k^2 \Psi=-4 \pi G a^{2} \bigg[\rho \Delta + \frac{\beta^2 \rho_ca^2}{k^2+ m^2a^2}  \rho_c \Delta_c + 3(\rho + p) \sigma \bigg] \ .
\label{eq:psi}
\ee

While not strictly relevant to {\tt MGCAMB}, let us note that combining Eqs. \eqref{eq:contic2} and \eqref{eq:Eulerc2} in the QSA limit, we obtain the second order differential equation for $\delta_c$:
\begin{equation}
\ddot{\delta}_c+ {\cal H} \dot{\delta}_c = -k^{2}\left[\Psi+\beta \delta \phi\right]
\label{eq:2ndorderc}
\end{equation}
which has the additional term on the right hand side, due to the fifth force mediated by the scalar field. The same equation for the baryons is
\begin{equation}
\ddot{\delta}_b+ {\cal H} \dot{\delta}_b = -k^{2}\Psi \ ,
\label{eq41}
\end{equation}
where we have omitted the baryon-photon coupling effect for the convenience of discussion. Hence, the total matter density contrast, $\delta = (\rho_c \delta_c + \rho_b \delta_b)/(\rho_c+\rho_b)$, obeys
\be
\ddot{\delta}+ {\cal H} \dot{\delta} = 4 \pi G a^2 \Big(1+\frac{2\tilde{\beta}^2k^2}{k^2+m^2a^2} \frac{\rho_c^2 \delta_c}{\rho^2 \delta}\Big)
\label{eq42}
\ee
where we have used Eq.~\eqref{eq:deltaphi2} to replace $\delta \phi$. This allows us to identify the effective gravitational coupling as
\be
G_{\text{eff}} =G \Big(1+\frac{2\tilde{\beta}^2k^2}{k^2+m^2a^2} \frac{\rho_c^2 \delta_c}{\rho^2 \delta}\Big) \ ,
\label{eq:mueff1}
\ee
which is very similar to the $G_{\text{eff}}$ on obtains from scalar-tensor theories with a universal coupling to matter,
\be
G_{\text{eff}} =G \Big(1+\frac{2\tilde{\beta}^2k^2}{k^2+m^2a^2} \Big) \ ,
\label{eq:mueff2}
\ee
making it challenging to distinguish between the two cases observationally~\cite{Bonvin:2022tii}.

\section{The synchronous gauge implementation of Einstein equations in the CDM-only coupled scalar field models} 
\label{appendixB}

While the contribution from the scalar field density perturbation to the Poisson equation (\ref{eq42}) is negligible for the QSA-compatible models that we study, for completeness, we show their implementation in {\tt MGCAMB}, which uses the synchronous gauge. Let us introduce
\be
C_\phi \equiv \frac{\beta^2 \rho_ca^2}{k^2+ m^2a^2}
\ee
and write
%

\ba
-{k^{2} \Psi \over 4 \pi G a^{2}}=\rho \Delta +C_{\phi} \rho_{c} \Delta_{c}+3(\rho+p) \sigma
\label{A-Poisson3} \\
{k^{2} (\Phi - \Psi) \over 4 \pi G a^{2}}= 3(\rho+p) \sigma \ ,
 \label{A-Poisson4}
\ea
where $\rho\Delta = \rho\delta + 3{\cal H}(\rho+p) \theta/k^2$ includes all species, including photons and neutrinos. Our goal is to derive the quantity $z = \dot{h}/(2k)$ appearing in {\tt CAMB}. Applying the transformation between the two gauges~\cite{Ma:1995ey},
\begin{align}
\Psi =\dot{\alpha}+{\cal H} \alpha \label{eq:gauge1} \\
\Phi=\eta -  {\cal H} \alpha 
\label{eq:gauge2}
\end{align}
where $\alpha =(\dot{h}+6 \dot{\eta}) / 2 k^{2}$, to \eqref{A-Poisson3} and \eqref{A-Poisson4}, allows us to write
\ba
\dot{\alpha}=-\eta-\frac{a^{2}}{2 k^{2}}\bigg[2\rho \Delta + C_{\phi} \rho_c \Delta_c+3 \rho\left(1+w\right) \sigma \bigg] 
\label{A-dotalpha} \\
\alpha = \frac{1}{H}\Bigg\{\eta+\frac{a^{2}}{2 k^{2}}\bigg[\rho \Delta+C_{\phi} \rho_c \Delta_c \bigg]\Bigg\} \ .
\label{A-alpha}
\ea
We can rewrite (\ref{A-alpha}) as
\begin{equation}
\eta={\cal H} \alpha-\frac{a^{2}}{2 k^{2}}[\rho \Delta+C_{\phi} \rho_c \Delta_c]={\cal H} \alpha -\frac{a^{2}}{2 k^{2}} \Gamma
\label{A-eta}
\end{equation}
where $\Gamma = \rho \Delta+C_{\phi}\rho_c \Delta_c$. We can also combine \eqref{A-dotalpha} and \eqref{A-alpha} to obtain
\be
\dot{\alpha}=-{\cal H} \alpha -\frac{a^{2}}{2 k^{2}}\big[\Gamma +3 \rho \left(1+w\right) \sigma \big] \ .
\label{A-dotalpha2}
\ee
Taking the derivative of \eqref{A-eta}, we obtain
\be
\dot{\eta}= \dot{{\cal H}} \alpha +{\cal H}\dot{\alpha}-\frac{a^{2}}{2 k^{2}} \dot{\Gamma}-\frac{a^{2}}{k^{2}} {\cal H} \Gamma 
\label{A-doteta1}
\ee
In order to calculate $\dot{\Gamma}$, we need to know $\dot{(\rho \Delta)}$. For standard matter, the conservation equations are 
\begin{align}
&\dot{\delta}=-(1+w)\left(\theta+\frac{\dot{h}}{2}\right)-3 {\cal H}\left(\frac{\delta p}{\delta \rho}-w\right) \delta \label{A-deltanormal}\\
&\dot{\theta}=-{\cal H}(1-3 w) \theta-\frac{\dot{w}}{1+w} \theta+\frac{\delta p / \delta \rho}{1+w} k^{2} \delta-k^{2} \sigma \ .
\label{A-thetanormal}
\end{align}
From this, we can write
\begin{align}
\dot{(\rho \Delta)}=&-3 {\cal H} \rho \Delta -(1+w) \rho \theta\bigg[1+\frac{3}{k^{2}}\big({{\cal H}}^{2}-\dot{{\cal H}}\big)\bigg]\nonumber \\
&-3 {\cal H}  \rho(1+w) \sigma-(1+w) \rho\left(k^{2} \alpha-3 \dot{\eta}\right)
\label{A-rhoDeltamdot}
\end{align}
For dark matter, the conservation equations are given by \eqref{eq:continuity-sync} and \eqref{eq:Eulerc3_QSA}, from which we find
\begin{align}
&\dot{(\rho_c \Delta_c)} = -3 {\cal H} \rho_c \Delta_c - \rho_c \theta_c \bigg[1+\frac{3}{k^{2}}\left({{\cal H}}^{2}-\dot{{\cal H}}\right)\bigg] \nonumber \\
&- \rho_c\left(k^{2} \alpha-3 \dot{\eta}\right)-(\beta \dot{\beta}+3{\cal H}\beta^2) \frac{\rho_c^2 (\delta_c - 3\alpha {\cal H})}{(k^2 +a^2 m^2)} \ ,
\label{A-rhoDeltadmdot}
\end{align}
where we have used $w=0$ and $\sigma = 0$ for CDM. We also have
\begin{equation}
\dot{\Gamma}=\dot{(\rho \Delta})+ C_{\phi} \dot{\left(\rho_c \Delta_c\right)}+\dot{C_{\phi}} \rho_c \Delta_c
\label{A-dotGamma1}
\end{equation}
Using Eqs.~\eqref{A-rhoDeltamdot} and \eqref{A-rhoDeltadmdot} in \eqref{A-dotGamma1}, we have
\begin{align}
&\dot{\Gamma}=-3 {\cal H} \rho \Delta-\rho \theta (1+w)\bigg[1+\frac{3}{k^{2}}\left({{\cal H}}^{2}-\dot{{\cal H}}\right)\bigg] \nonumber \\
&-\rho(1+w)\left(k^{2} \alpha-3 \dot{\eta}\right) -3 {\cal H} \rho \left(1+w\right) \sigma\nonumber \\
&+C_{\phi} \bigg\{
-3 H \rho_c \Delta_c - \rho_c \theta_c \bigg[1+\frac{3}{k^{2}}\left({{\cal H}}^{2}-\dot{{\cal H}}\right)\bigg] \nonumber \\
&- \rho_c\left(k^{2} \alpha-3 \dot{\eta}\right)-(\beta \dot{\beta}+3{\cal H}\beta^2) \frac{\rho_c^2 (\delta_c - 3\alpha {\cal H})}{(k^2 +a^2 m^2)}
\bigg\} \nonumber\\ 
&+ \dot{C_{\phi}}\rho_c \Delta_c \ .
\label{A-dotGamma2}
\end{align} 
We then substitute $\Gamma$ and $\dot{\Gamma}$ into the equation for $\dot{\eta}$, and solve for $\dot{\eta}$, to find
\begin{align}
\dot{\eta}&=\frac{1}{2} \frac{a^{2}}{k^{2}+\frac{3}{2} a^{2}[\rho(1+w)+C_{\phi}\rho_c]}\nonumber \\
&\Bigg\{k \rho q\Big[1+\frac{3({{\cal H}}^2-\dot{{\cal H}})}{k^{2}}\Big] \nonumber \\
&-\rho_c \Delta_c \dot{C_{\phi}} +k^{2} \alpha\Big[C_{\phi} \rho_c-\rho_{\mathrm{DE}}\left(1+w_{\mathrm{DE}}\right)\Big] \nonumber \\
&+ (1+C_{\phi})(\beta \dot{\beta}+3{\cal H}\beta^2) \frac{\rho_c^2 (\delta_c - 3\alpha {\cal H})}{k^2/a^2 + m^2}\Bigg\} \ ,
 \label{A-doteta2}
\end{align}
with $k q = (1+w) \theta$. Finally, we can obtain $z$ used in the code from
\begin{equation}
 z = k \alpha - 3 \frac{\dot{\eta}}{k} \ .
 	\label{eq:z2}
\end{equation}


\begin{thebibliography}{94}
  \expandafter\ifx\csname natexlab\endcsname\relax\def\natexlab#1{#1}\fi
  \expandafter\ifx\csname bibnamefont\endcsname\relax
    \def\bibnamefont#1{#1}\fi
  \expandafter\ifx\csname bibfnamefont\endcsname\relax
    \def\bibfnamefont#1{#1}\fi
  \expandafter\ifx\csname citenamefont\endcsname\relax
    \def\citenamefont#1{#1}\fi
  \expandafter\ifx\csname url\endcsname\relax
    \def\url#1{\texttt{#1}}\fi
  \expandafter\ifx\csname urlprefix\endcsname\relax\def\urlprefix{URL }\fi
  \providecommand{\bibinfo}[2]{#2}
  \providecommand{\eprint}[2][]{\url{#2}}
  
  \bibitem[{\citenamefont{Perlmutter et~al.}(1999)}]{Perlmutter:1998np}
  \bibinfo{author}{\bibfnamefont{S.}~\bibnamefont{Perlmutter}}
    \bibnamefont{et~al.} (\bibinfo{collaboration}{Supernova Cosmology Project}),
    \bibinfo{journal}{Astrophys. J.} \textbf{\bibinfo{volume}{517}},
    \bibinfo{pages}{565} (\bibinfo{year}{1999}), \eprint{astro-ph/9812133}.
  
  \bibitem[{\citenamefont{Riess et~al.}(1998)}]{Riess:1998cb}
  \bibinfo{author}{\bibfnamefont{A.~G.} \bibnamefont{Riess}} \bibnamefont{et~al.}
    (\bibinfo{collaboration}{Supernova Search Team}), \bibinfo{journal}{Astron.
    J.} \textbf{\bibinfo{volume}{116}}, \bibinfo{pages}{1009}
    (\bibinfo{year}{1998}), \eprint{astro-ph/9805201}.
  
  \bibitem[{\citenamefont{Weinberg}(1989)}]{Weinberg:1988cp}
  \bibinfo{author}{\bibfnamefont{S.}~\bibnamefont{Weinberg}},
    \bibinfo{journal}{Rev. Mod. Phys.} \textbf{\bibinfo{volume}{61}},
    \bibinfo{pages}{1} (\bibinfo{year}{1989}).
  
  \bibitem[{\citenamefont{Burgess}(2015)}]{Burgess:2013ara}
  \bibinfo{author}{\bibfnamefont{C.~P.} \bibnamefont{Burgess}}, in
    \emph{\bibinfo{booktitle}{{100e Ecole d'Ete de Physique: Post-Planck
    Cosmology Les Houches, France, July 8-August 2, 2013}}}
    (\bibinfo{year}{2015}), pp. \bibinfo{pages}{149--197}, \eprint{1309.4133},
    \urlprefix\url{http://inspirehep.net/record/1254422/files/arXiv:1309.4133.pdf}.
  
  \bibitem[{\citenamefont{Silvestri and Trodden}(2009)}]{Silvestri:2009hh}
  \bibinfo{author}{\bibfnamefont{A.}~\bibnamefont{Silvestri}} \bibnamefont{and}
    \bibinfo{author}{\bibfnamefont{M.}~\bibnamefont{Trodden}},
    \bibinfo{journal}{Rept. Prog. Phys.} \textbf{\bibinfo{volume}{72}},
    \bibinfo{pages}{096901} (\bibinfo{year}{2009}), \eprint{0904.0024}.
  
  \bibitem[{\citenamefont{Clifton et~al.}(2012)\citenamefont{Clifton, Ferreira,
    Padilla, and Skordis}}]{Clifton:2011jh}
  \bibinfo{author}{\bibfnamefont{T.}~\bibnamefont{Clifton}},
    \bibinfo{author}{\bibfnamefont{P.~G.} \bibnamefont{Ferreira}},
    \bibinfo{author}{\bibfnamefont{A.}~\bibnamefont{Padilla}}, \bibnamefont{and}
    \bibinfo{author}{\bibfnamefont{C.}~\bibnamefont{Skordis}},
    \bibinfo{journal}{Phys. Rept.} \textbf{\bibinfo{volume}{513}},
    \bibinfo{pages}{1} (\bibinfo{year}{2012}), \eprint{1106.2476}.
  
  \bibitem[{\citenamefont{Joyce et~al.}(2015)\citenamefont{Joyce, Jain, Khoury,
    and Trodden}}]{Joyce:2014kja}
  \bibinfo{author}{\bibfnamefont{A.}~\bibnamefont{Joyce}},
    \bibinfo{author}{\bibfnamefont{B.}~\bibnamefont{Jain}},
    \bibinfo{author}{\bibfnamefont{J.}~\bibnamefont{Khoury}}, \bibnamefont{and}
    \bibinfo{author}{\bibfnamefont{M.}~\bibnamefont{Trodden}},
    \bibinfo{journal}{Phys. Rept.} \textbf{\bibinfo{volume}{568}},
    \bibinfo{pages}{1} (\bibinfo{year}{2015}), \eprint{1407.0059}.
  
  \bibitem[{\citenamefont{Koyama}(2016)}]{Koyama:2015vza}
  \bibinfo{author}{\bibfnamefont{K.}~\bibnamefont{Koyama}},
    \bibinfo{journal}{Rept. Prog. Phys.} \textbf{\bibinfo{volume}{79}},
    \bibinfo{pages}{046902} (\bibinfo{year}{2016}), \eprint{1504.04623}.
  
  \bibitem[{\citenamefont{Ishak}(2019)}]{Ishak:2018his}
  \bibinfo{author}{\bibfnamefont{M.}~\bibnamefont{Ishak}},
    \bibinfo{journal}{Living Rev. Rel.} \textbf{\bibinfo{volume}{22}},
    \bibinfo{pages}{1} (\bibinfo{year}{2019}), \eprint{1806.10122}.
  
  \bibitem[{\citenamefont{Abdalla et~al.}(2022)}]{Abdalla:2022yfr}
  \bibinfo{author}{\bibfnamefont{E.}~\bibnamefont{Abdalla}} \bibnamefont{et~al.},
    \bibinfo{journal}{JHEAp} \textbf{\bibinfo{volume}{34}}, \bibinfo{pages}{49}
    (\bibinfo{year}{2022}), \eprint{2203.06142}.
  
  \bibitem[{\citenamefont{Zhao et~al.}(2009)\citenamefont{Zhao, Pogosian,
    Silvestri, and Zylberberg}}]{Zhao:2008bn}
  \bibinfo{author}{\bibfnamefont{G.-B.} \bibnamefont{Zhao}},
    \bibinfo{author}{\bibfnamefont{L.}~\bibnamefont{Pogosian}},
    \bibinfo{author}{\bibfnamefont{A.}~\bibnamefont{Silvestri}},
    \bibnamefont{and}
    \bibinfo{author}{\bibfnamefont{J.}~\bibnamefont{Zylberberg}},
    \bibinfo{journal}{Phys. Rev.} \textbf{\bibinfo{volume}{D79}},
    \bibinfo{pages}{083513} (\bibinfo{year}{2009}), \eprint{0809.3791}.
  
  \bibitem[{\citenamefont{Hojjati et~al.}(2011)\citenamefont{Hojjati, Pogosian,
    and Zhao}}]{Hojjati:2011ix}
  \bibinfo{author}{\bibfnamefont{A.}~\bibnamefont{Hojjati}},
    \bibinfo{author}{\bibfnamefont{L.}~\bibnamefont{Pogosian}}, \bibnamefont{and}
    \bibinfo{author}{\bibfnamefont{G.-B.} \bibnamefont{Zhao}},
    \bibinfo{journal}{JCAP} \textbf{\bibinfo{volume}{1108}}, \bibinfo{pages}{005}
    (\bibinfo{year}{2011}), \eprint{1106.4543}.
  
  \bibitem[{\citenamefont{Zucca et~al.}(2019)\citenamefont{Zucca, Pogosian,
    Silvestri, and Zhao}}]{Zucca:2019xhg}
  \bibinfo{author}{\bibfnamefont{A.}~\bibnamefont{Zucca}},
    \bibinfo{author}{\bibfnamefont{L.}~\bibnamefont{Pogosian}},
    \bibinfo{author}{\bibfnamefont{A.}~\bibnamefont{Silvestri}},
    \bibnamefont{and} \bibinfo{author}{\bibfnamefont{G.-B.} \bibnamefont{Zhao}},
    \bibinfo{journal}{JCAP} \textbf{\bibinfo{volume}{05}}, \bibinfo{pages}{001}
    (\bibinfo{year}{2019}), \eprint{1901.05956}.
  
  \bibitem[{\citenamefont{Lewis et~al.}(2000)\citenamefont{Lewis, Challinor, and
    Lasenby}}]{Lewis:1999bs}
  \bibinfo{author}{\bibfnamefont{A.}~\bibnamefont{Lewis}},
    \bibinfo{author}{\bibfnamefont{A.}~\bibnamefont{Challinor}},
    \bibnamefont{and} \bibinfo{author}{\bibfnamefont{A.}~\bibnamefont{Lasenby}},
    \bibinfo{journal}{Astrophys. J.} \textbf{\bibinfo{volume}{538}},
    \bibinfo{pages}{473} (\bibinfo{year}{2000}), \eprint{astro-ph/9911177}.
  
  \bibitem[{cam()}]{camb}
  \bibinfo{howpublished}{\url{http://camb.info}}.
  
  \bibitem[{\citenamefont{Lewis and Bridle}(2002)}]{Lewis:2002ah}
  \bibinfo{author}{\bibfnamefont{A.}~\bibnamefont{Lewis}} \bibnamefont{and}
    \bibinfo{author}{\bibfnamefont{S.}~\bibnamefont{Bridle}},
    \bibinfo{journal}{Phys. Rev.} \textbf{\bibinfo{volume}{D66}},
    \bibinfo{pages}{103511} (\bibinfo{year}{2002}), \eprint{astro-ph/0205436}.
  
  \bibitem[{cos()}]{cosmomc}
  \bibinfo{howpublished}{\url{http://cosmologist.info/cosmomc}}.
  
  \bibitem[{\citenamefont{Simpson et~al.}(2013)}]{Simpson:2012ra}
  \bibinfo{author}{\bibfnamefont{F.}~\bibnamefont{Simpson}} \bibnamefont{et~al.},
    \bibinfo{journal}{Mon. Not. Roy. Astron. Soc.}
    \textbf{\bibinfo{volume}{429}}, \bibinfo{pages}{2249} (\bibinfo{year}{2013}),
    \eprint{1212.3339}.
  
  \bibitem[{\citenamefont{Ade et~al.}(2016)}]{Ade:2015rim}
  \bibinfo{author}{\bibfnamefont{P.~A.~R.} \bibnamefont{Ade}}
    \bibnamefont{et~al.} (\bibinfo{collaboration}{Planck}),
    \bibinfo{journal}{Astron. Astrophys.} \textbf{\bibinfo{volume}{594}},
    \bibinfo{pages}{A14} (\bibinfo{year}{2016}), \eprint{1502.01590}.
  
  \bibitem[{\citenamefont{Abbott et~al.}(2019)}]{DES:2018ufa}
  \bibinfo{author}{\bibfnamefont{T.~M.~C.} \bibnamefont{Abbott}}
    \bibnamefont{et~al.} (\bibinfo{collaboration}{DES}), \bibinfo{journal}{Phys.
    Rev. D} \textbf{\bibinfo{volume}{99}}, \bibinfo{pages}{123505}
    (\bibinfo{year}{2019}), \eprint{1810.02499}.
  
  \bibitem[{\citenamefont{Capozziello et~al.}(2003)\citenamefont{Capozziello,
    Carloni, and Troisi}}]{Capozziello:2003tk}
  \bibinfo{author}{\bibfnamefont{S.}~\bibnamefont{Capozziello}},
    \bibinfo{author}{\bibfnamefont{S.}~\bibnamefont{Carloni}}, \bibnamefont{and}
    \bibinfo{author}{\bibfnamefont{A.}~\bibnamefont{Troisi}},
    \bibinfo{journal}{Recent Res. Dev. Astron. Astrophys.}
    \textbf{\bibinfo{volume}{1}}, \bibinfo{pages}{625} (\bibinfo{year}{2003}),
    \eprint{astro-ph/0303041}.
  
  \bibitem[{\citenamefont{Carroll et~al.}(2004)\citenamefont{Carroll, Duvvuri,
    Trodden, and Turner}}]{Carroll:2003wy}
  \bibinfo{author}{\bibfnamefont{S.~M.} \bibnamefont{Carroll}},
    \bibinfo{author}{\bibfnamefont{V.}~\bibnamefont{Duvvuri}},
    \bibinfo{author}{\bibfnamefont{M.}~\bibnamefont{Trodden}}, \bibnamefont{and}
    \bibinfo{author}{\bibfnamefont{M.~S.} \bibnamefont{Turner}},
    \bibinfo{journal}{Phys. Rev.} \textbf{\bibinfo{volume}{D70}},
    \bibinfo{pages}{043528} (\bibinfo{year}{2004}), \eprint{astro-ph/0306438}.
  
  \bibitem[{\citenamefont{Hu and Sawicki}(2007)}]{Hu:2007nk}
  \bibinfo{author}{\bibfnamefont{W.}~\bibnamefont{Hu}} \bibnamefont{and}
    \bibinfo{author}{\bibfnamefont{I.}~\bibnamefont{Sawicki}},
    \bibinfo{journal}{Phys. Rev.} \textbf{\bibinfo{volume}{D76}},
    \bibinfo{pages}{064004} (\bibinfo{year}{2007}), \eprint{0705.1158}.
  
  \bibitem[{\citenamefont{Pogosian and Silvestri}(2008)}]{Pogosian:2007sw}
  \bibinfo{author}{\bibfnamefont{L.}~\bibnamefont{Pogosian}} \bibnamefont{and}
    \bibinfo{author}{\bibfnamefont{A.}~\bibnamefont{Silvestri}},
    \bibinfo{journal}{Phys. Rev. D} \textbf{\bibinfo{volume}{77}},
    \bibinfo{pages}{023503} (\bibinfo{year}{2008}), \bibinfo{note}{[Erratum:
    Phys.Rev.D 81, 049901 (2010)]}, \eprint{0709.0296}.
  
  \bibitem[{\citenamefont{Dvali et~al.}(2000)\citenamefont{Dvali, Gabadadze, and
    Porrati}}]{Dvali:2000rv}
  \bibinfo{author}{\bibfnamefont{G.~R.} \bibnamefont{Dvali}},
    \bibinfo{author}{\bibfnamefont{G.}~\bibnamefont{Gabadadze}},
    \bibnamefont{and} \bibinfo{author}{\bibfnamefont{M.}~\bibnamefont{Porrati}},
    \bibinfo{journal}{Phys. Lett. B} \textbf{\bibinfo{volume}{484}},
    \bibinfo{pages}{112} (\bibinfo{year}{2000}), \eprint{hep-th/0002190}.
  
  \bibitem[{\citenamefont{Koyama and Maartens}(2006)}]{Koyama:2005kd}
  \bibinfo{author}{\bibfnamefont{K.}~\bibnamefont{Koyama}} \bibnamefont{and}
    \bibinfo{author}{\bibfnamefont{R.}~\bibnamefont{Maartens}},
    \bibinfo{journal}{JCAP} \textbf{\bibinfo{volume}{01}}, \bibinfo{pages}{016}
    (\bibinfo{year}{2006}), \eprint{astro-ph/0511634}.
  
  \bibitem[{\citenamefont{Gubitosi et~al.}(2013)\citenamefont{Gubitosi, Piazza,
    and Vernizzi}}]{Gubitosi:2012hu}
  \bibinfo{author}{\bibfnamefont{G.}~\bibnamefont{Gubitosi}},
    \bibinfo{author}{\bibfnamefont{F.}~\bibnamefont{Piazza}}, \bibnamefont{and}
    \bibinfo{author}{\bibfnamefont{F.}~\bibnamefont{Vernizzi}},
    \bibinfo{journal}{JCAP} \textbf{\bibinfo{volume}{1302}}, \bibinfo{pages}{032}
    (\bibinfo{year}{2013}), \bibinfo{note}{[JCAP1302,032(2013)]},
    \eprint{1210.0201}.
  
  \bibitem[{\citenamefont{Bloomfield et~al.}(2013)\citenamefont{Bloomfield,
    Flanagan, Park, and Watson}}]{Bloomfield:2012ff}
  \bibinfo{author}{\bibfnamefont{J.~K.} \bibnamefont{Bloomfield}},
    \bibinfo{author}{\bibfnamefont{E.~E.} \bibnamefont{Flanagan}},
    \bibinfo{author}{\bibfnamefont{M.}~\bibnamefont{Park}}, \bibnamefont{and}
    \bibinfo{author}{\bibfnamefont{S.}~\bibnamefont{Watson}},
    \bibinfo{journal}{JCAP} \textbf{\bibinfo{volume}{1308}}, \bibinfo{pages}{010}
    (\bibinfo{year}{2013}), \eprint{1211.7054}.
  
  \bibitem[{\citenamefont{Gleyzes et~al.}(2013)\citenamefont{Gleyzes, Langlois,
    Piazza, and Vernizzi}}]{Gleyzes:2013ooa}
  \bibinfo{author}{\bibfnamefont{J.}~\bibnamefont{Gleyzes}},
    \bibinfo{author}{\bibfnamefont{D.}~\bibnamefont{Langlois}},
    \bibinfo{author}{\bibfnamefont{F.}~\bibnamefont{Piazza}}, \bibnamefont{and}
    \bibinfo{author}{\bibfnamefont{F.}~\bibnamefont{Vernizzi}},
    \bibinfo{journal}{JCAP} \textbf{\bibinfo{volume}{1308}}, \bibinfo{pages}{025}
    (\bibinfo{year}{2013}), \eprint{1304.4840}.
  
  \bibitem[{\citenamefont{Bloomfield}(2013)}]{Bloomfield:2013efa}
  \bibinfo{author}{\bibfnamefont{J.}~\bibnamefont{Bloomfield}},
    \bibinfo{journal}{JCAP} \textbf{\bibinfo{volume}{1312}}, \bibinfo{pages}{044}
    (\bibinfo{year}{2013}), \eprint{1304.6712}.
  
  \bibitem[{\citenamefont{Silvestri et~al.}(2013)\citenamefont{Silvestri,
    Pogosian, and Buniy}}]{Silvestri:2013ne}
  \bibinfo{author}{\bibfnamefont{A.}~\bibnamefont{Silvestri}},
    \bibinfo{author}{\bibfnamefont{L.}~\bibnamefont{Pogosian}}, \bibnamefont{and}
    \bibinfo{author}{\bibfnamefont{R.~V.} \bibnamefont{Buniy}},
    \bibinfo{journal}{Phys. Rev.} \textbf{\bibinfo{volume}{D87}},
    \bibinfo{pages}{104015} (\bibinfo{year}{2013}), \eprint{1302.1193}.
  
  \bibitem[{\citenamefont{Bellini and Sawicki}(2014)}]{Bellini:2014fua}
  \bibinfo{author}{\bibfnamefont{E.}~\bibnamefont{Bellini}} \bibnamefont{and}
    \bibinfo{author}{\bibfnamefont{I.}~\bibnamefont{Sawicki}},
    \bibinfo{journal}{JCAP} \textbf{\bibinfo{volume}{1407}}, \bibinfo{pages}{050}
    (\bibinfo{year}{2014}), \eprint{1404.3713}.
  
  \bibitem[{\citenamefont{Gleyzes
    et~al.}(2015{\natexlab{a}})\citenamefont{Gleyzes, Langlois, and
    Vernizzi}}]{Gleyzes:2014rba}
  \bibinfo{author}{\bibfnamefont{J.}~\bibnamefont{Gleyzes}},
    \bibinfo{author}{\bibfnamefont{D.}~\bibnamefont{Langlois}}, \bibnamefont{and}
    \bibinfo{author}{\bibfnamefont{F.}~\bibnamefont{Vernizzi}},
    \bibinfo{journal}{Int. J. Mod. Phys.} \textbf{\bibinfo{volume}{D23}},
    \bibinfo{pages}{1443010} (\bibinfo{year}{2015}{\natexlab{a}}),
    \eprint{1411.3712}.
  
  \bibitem[{\citenamefont{Gleyzes
    et~al.}(2015{\natexlab{b}})\citenamefont{Gleyzes, Langlois, Mancarella, and
    Vernizzi}}]{Gleyzes:2015pma}
  \bibinfo{author}{\bibfnamefont{J.}~\bibnamefont{Gleyzes}},
    \bibinfo{author}{\bibfnamefont{D.}~\bibnamefont{Langlois}},
    \bibinfo{author}{\bibfnamefont{M.}~\bibnamefont{Mancarella}},
    \bibnamefont{and} \bibinfo{author}{\bibfnamefont{F.}~\bibnamefont{Vernizzi}},
    \bibinfo{journal}{JCAP} \textbf{\bibinfo{volume}{1508}}, \bibinfo{pages}{054}
    (\bibinfo{year}{2015}{\natexlab{b}}), \eprint{1504.05481}.
  
  \bibitem[{\citenamefont{Horndeski}(1974)}]{Horndeski:1974wa}
  \bibinfo{author}{\bibfnamefont{G.~W.} \bibnamefont{Horndeski}},
    \bibinfo{journal}{Int. J. Theor. Phys.} \textbf{\bibinfo{volume}{10}},
    \bibinfo{pages}{363} (\bibinfo{year}{1974}).
  
  \bibitem[{\citenamefont{Deffayet et~al.}(2011)\citenamefont{Deffayet, Gao,
    Steer, and Zahariade}}]{Deffayet:2011gz}
  \bibinfo{author}{\bibfnamefont{C.}~\bibnamefont{Deffayet}},
    \bibinfo{author}{\bibfnamefont{X.}~\bibnamefont{Gao}},
    \bibinfo{author}{\bibfnamefont{D.~A.} \bibnamefont{Steer}}, \bibnamefont{and}
    \bibinfo{author}{\bibfnamefont{G.}~\bibnamefont{Zahariade}},
    \bibinfo{journal}{Phys. Rev.} \textbf{\bibinfo{volume}{D84}},
    \bibinfo{pages}{064039} (\bibinfo{year}{2011}), \eprint{1103.3260}.
  
  \bibitem[{\citenamefont{Gleyzes
    et~al.}(2015{\natexlab{c}})\citenamefont{Gleyzes, Langlois, Piazza, and
    Vernizzi}}]{Gleyzes:2014qga}
  \bibinfo{author}{\bibfnamefont{J.}~\bibnamefont{Gleyzes}},
    \bibinfo{author}{\bibfnamefont{D.}~\bibnamefont{Langlois}},
    \bibinfo{author}{\bibfnamefont{F.}~\bibnamefont{Piazza}}, \bibnamefont{and}
    \bibinfo{author}{\bibfnamefont{F.}~\bibnamefont{Vernizzi}},
    \bibinfo{journal}{JCAP} \textbf{\bibinfo{volume}{1502}}, \bibinfo{pages}{018}
    (\bibinfo{year}{2015}{\natexlab{c}}), \eprint{1408.1952}.
  
  \bibitem[{\citenamefont{Gleyzes
    et~al.}(2015{\natexlab{d}})\citenamefont{Gleyzes, Langlois, Piazza, and
    Vernizzi}}]{Gleyzes:2014dya}
  \bibinfo{author}{\bibfnamefont{J.}~\bibnamefont{Gleyzes}},
    \bibinfo{author}{\bibfnamefont{D.}~\bibnamefont{Langlois}},
    \bibinfo{author}{\bibfnamefont{F.}~\bibnamefont{Piazza}}, \bibnamefont{and}
    \bibinfo{author}{\bibfnamefont{F.}~\bibnamefont{Vernizzi}},
    \bibinfo{journal}{Phys. Rev. Lett.} \textbf{\bibinfo{volume}{114}},
    \bibinfo{pages}{211101} (\bibinfo{year}{2015}{\natexlab{d}}),
    \eprint{1404.6495}.
  
  \bibitem[{\citenamefont{Langlois and Noui}(2016)}]{Langlois:2015cwa}
  \bibinfo{author}{\bibfnamefont{D.}~\bibnamefont{Langlois}} \bibnamefont{and}
    \bibinfo{author}{\bibfnamefont{K.}~\bibnamefont{Noui}},
    \bibinfo{journal}{JCAP} \textbf{\bibinfo{volume}{02}}, \bibinfo{pages}{034}
    (\bibinfo{year}{2016}), \eprint{1510.06930}.
  
  \bibitem[{\citenamefont{Ben~Achour et~al.}(2016)\citenamefont{Ben~Achour,
    Langlois, and Noui}}]{BenAchour:2016cay}
  \bibinfo{author}{\bibfnamefont{J.}~\bibnamefont{Ben~Achour}},
    \bibinfo{author}{\bibfnamefont{D.}~\bibnamefont{Langlois}}, \bibnamefont{and}
    \bibinfo{author}{\bibfnamefont{K.}~\bibnamefont{Noui}},
    \bibinfo{journal}{Phys. Rev. D} \textbf{\bibinfo{volume}{93}},
    \bibinfo{pages}{124005} (\bibinfo{year}{2016}), \eprint{1602.08398}.
  
  \bibitem[{\citenamefont{Hu et~al.}(2014)\citenamefont{Hu, Raveri, Frusciante,
    and Silvestri}}]{Hu:2013twa}
  \bibinfo{author}{\bibfnamefont{B.}~\bibnamefont{Hu}},
    \bibinfo{author}{\bibfnamefont{M.}~\bibnamefont{Raveri}},
    \bibinfo{author}{\bibfnamefont{N.}~\bibnamefont{Frusciante}},
    \bibnamefont{and}
    \bibinfo{author}{\bibfnamefont{A.}~\bibnamefont{Silvestri}},
    \bibinfo{journal}{Phys. Rev. D} \textbf{\bibinfo{volume}{89}},
    \bibinfo{pages}{103530} (\bibinfo{year}{2014}), \eprint{1312.5742}.
  
  \bibitem[{\citenamefont{{Hu} et~al.}(2014)\citenamefont{{Hu}, {Raveri},
    {Frusciante}, and {Silvestri}}}]{2014arXiv1405.3590H}
  \bibinfo{author}{\bibfnamefont{B.}~\bibnamefont{{Hu}}},
    \bibinfo{author}{\bibfnamefont{M.}~\bibnamefont{{Raveri}}},
    \bibinfo{author}{\bibfnamefont{N.}~\bibnamefont{{Frusciante}}},
    \bibnamefont{and}
    \bibinfo{author}{\bibfnamefont{A.}~\bibnamefont{{Silvestri}}},
    \bibinfo{journal}{ArXiv e-prints}  (\bibinfo{year}{2014}),
    \eprint{1405.3590}.
  
  \bibitem[{\citenamefont{Raveri et~al.}(2014)\citenamefont{Raveri, Hu,
    Frusciante, and Silvestri}}]{Raveri:2014cka}
  \bibinfo{author}{\bibfnamefont{M.}~\bibnamefont{Raveri}},
    \bibinfo{author}{\bibfnamefont{B.}~\bibnamefont{Hu}},
    \bibinfo{author}{\bibfnamefont{N.}~\bibnamefont{Frusciante}},
    \bibnamefont{and}
    \bibinfo{author}{\bibfnamefont{A.}~\bibnamefont{Silvestri}},
    \bibinfo{journal}{Phys. Rev.} \textbf{\bibinfo{volume}{D90}},
    \bibinfo{pages}{043513} (\bibinfo{year}{2014}), \eprint{1405.1022}.
  
  \bibitem[{\citenamefont{Zumalacarregui
    et~al.}(2016)\citenamefont{Zumalacarregui, Bellini, Sawicki, and
    Lesgourgues}}]{Zumalacarregui:2016pph}
  \bibinfo{author}{\bibfnamefont{M.}~\bibnamefont{Zumalacarregui}},
    \bibinfo{author}{\bibfnamefont{E.}~\bibnamefont{Bellini}},
    \bibinfo{author}{\bibfnamefont{I.}~\bibnamefont{Sawicki}}, \bibnamefont{and}
    \bibinfo{author}{\bibfnamefont{J.}~\bibnamefont{Lesgourgues}}
    (\bibinfo{year}{2016}), \eprint{1605.06102}.
  
  \bibitem[{\citenamefont{Bellini et~al.}(2020)\citenamefont{Bellini, Sawicki,
    and Zumalac\'arregui}}]{Bellini:2019syt}
  \bibinfo{author}{\bibfnamefont{E.}~\bibnamefont{Bellini}},
    \bibinfo{author}{\bibfnamefont{I.}~\bibnamefont{Sawicki}}, \bibnamefont{and}
    \bibinfo{author}{\bibfnamefont{M.}~\bibnamefont{Zumalac\'arregui}},
    \bibinfo{journal}{JCAP} \textbf{\bibinfo{volume}{02}}, \bibinfo{pages}{008}
    (\bibinfo{year}{2020}), \eprint{1909.01828}.
  
  \bibitem[{\citenamefont{Sakr and Martinelli}(2022)}]{Sakr:2021ylx}
  \bibinfo{author}{\bibfnamefont{Z.}~\bibnamefont{Sakr}} \bibnamefont{and}
    \bibinfo{author}{\bibfnamefont{M.}~\bibnamefont{Martinelli}},
    \bibinfo{journal}{JCAP} \textbf{\bibinfo{volume}{05}}, \bibinfo{pages}{030}
    (\bibinfo{year}{2022}), \eprint{2112.14175}.
  
  \bibitem[{\citenamefont{Dossett et~al.}(2011)\citenamefont{Dossett, Ishak, and
    Moldenhauer}}]{Dossett:2011tn}
  \bibinfo{author}{\bibfnamefont{J.~N.} \bibnamefont{Dossett}},
    \bibinfo{author}{\bibfnamefont{M.}~\bibnamefont{Ishak}}, \bibnamefont{and}
    \bibinfo{author}{\bibfnamefont{J.}~\bibnamefont{Moldenhauer}},
    \bibinfo{journal}{Phys. Rev. D} \textbf{\bibinfo{volume}{84}},
    \bibinfo{pages}{123001} (\bibinfo{year}{2011}), \eprint{1109.4583}.
  
  \bibitem[{\citenamefont{Garcia-Quintero
    et~al.}(2019)\citenamefont{Garcia-Quintero, Ishak, Fox, and
    Dossett}}]{Garcia-Quintero:2019xal}
  \bibinfo{author}{\bibfnamefont{C.}~\bibnamefont{Garcia-Quintero}},
    \bibinfo{author}{\bibfnamefont{M.}~\bibnamefont{Ishak}},
    \bibinfo{author}{\bibfnamefont{L.}~\bibnamefont{Fox}}, \bibnamefont{and}
    \bibinfo{author}{\bibfnamefont{J.}~\bibnamefont{Dossett}},
    \bibinfo{journal}{Phys. Rev. D} \textbf{\bibinfo{volume}{100}},
    \bibinfo{pages}{103530} (\bibinfo{year}{2019}), \eprint{1908.00290}.
  
  \bibitem[{\citenamefont{Garcia-Quintero and
    Ishak}(2021)}]{Garcia-Quintero:2020mja}
  \bibinfo{author}{\bibfnamefont{C.}~\bibnamefont{Garcia-Quintero}}
    \bibnamefont{and} \bibinfo{author}{\bibfnamefont{M.}~\bibnamefont{Ishak}},
    \bibinfo{journal}{Mon. Not. Roy. Astron. Soc.}
    \textbf{\bibinfo{volume}{506}}, \bibinfo{pages}{1704} (\bibinfo{year}{2021}),
    \eprint{2009.01189}.
  
  \bibitem[{\citenamefont{Espejo et~al.}(2019)\citenamefont{Espejo, Peirone,
    Raveri, Koyama, Pogosian, and Silvestri}}]{Espejo:2018hxa}
  \bibinfo{author}{\bibfnamefont{J.}~\bibnamefont{Espejo}},
    \bibinfo{author}{\bibfnamefont{S.}~\bibnamefont{Peirone}},
    \bibinfo{author}{\bibfnamefont{M.}~\bibnamefont{Raveri}},
    \bibinfo{author}{\bibfnamefont{K.}~\bibnamefont{Koyama}},
    \bibinfo{author}{\bibfnamefont{L.}~\bibnamefont{Pogosian}}, \bibnamefont{and}
    \bibinfo{author}{\bibfnamefont{A.}~\bibnamefont{Silvestri}},
    \bibinfo{journal}{Phys. Rev. D} \textbf{\bibinfo{volume}{99}},
    \bibinfo{pages}{023512} (\bibinfo{year}{2019}), \eprint{1809.01121}.
  
  \bibitem[{\citenamefont{Pogosian and Silvestri}(2016)}]{Pogosian:2016pwr}
  \bibinfo{author}{\bibfnamefont{L.}~\bibnamefont{Pogosian}} \bibnamefont{and}
    \bibinfo{author}{\bibfnamefont{A.}~\bibnamefont{Silvestri}},
    \bibinfo{journal}{Phys. Rev.} \textbf{\bibinfo{volume}{D94}},
    \bibinfo{pages}{104014} (\bibinfo{year}{2016}), \eprint{1606.05339}.
  
  \bibitem[{\citenamefont{Pogosian et~al.}(2022)\citenamefont{Pogosian, Raveri,
    Koyama, Martinelli, Silvestri, Zhao, Li, Peirone, and
    Zucca}}]{Pogosian:2021mcs}
  \bibinfo{author}{\bibfnamefont{L.}~\bibnamefont{Pogosian}},
    \bibinfo{author}{\bibfnamefont{M.}~\bibnamefont{Raveri}},
    \bibinfo{author}{\bibfnamefont{K.}~\bibnamefont{Koyama}},
    \bibinfo{author}{\bibfnamefont{M.}~\bibnamefont{Martinelli}},
    \bibinfo{author}{\bibfnamefont{A.}~\bibnamefont{Silvestri}},
    \bibinfo{author}{\bibfnamefont{G.-B.} \bibnamefont{Zhao}},
    \bibinfo{author}{\bibfnamefont{J.}~\bibnamefont{Li}},
    \bibinfo{author}{\bibfnamefont{S.}~\bibnamefont{Peirone}}, \bibnamefont{and}
    \bibinfo{author}{\bibfnamefont{A.}~\bibnamefont{Zucca}},
    \bibinfo{journal}{Nature Astron.} \textbf{\bibinfo{volume}{6}},
    \bibinfo{pages}{1484} (\bibinfo{year}{2022}), \eprint{2107.12992}.
  
  \bibitem[{\citenamefont{Raveri et~al.}(2021)\citenamefont{Raveri, Pogosian,
    Martinelli, Koyama, Silvestri, Zhao, Li, Peirone, and
    Zucca}}]{Raveri:2021dbu}
  \bibinfo{author}{\bibfnamefont{M.}~\bibnamefont{Raveri}},
    \bibinfo{author}{\bibfnamefont{L.}~\bibnamefont{Pogosian}},
    \bibinfo{author}{\bibfnamefont{M.}~\bibnamefont{Martinelli}},
    \bibinfo{author}{\bibfnamefont{K.}~\bibnamefont{Koyama}},
    \bibinfo{author}{\bibfnamefont{A.}~\bibnamefont{Silvestri}},
    \bibinfo{author}{\bibfnamefont{G.-B.} \bibnamefont{Zhao}},
    \bibinfo{author}{\bibfnamefont{J.}~\bibnamefont{Li}},
    \bibinfo{author}{\bibfnamefont{S.}~\bibnamefont{Peirone}}, \bibnamefont{and}
    \bibinfo{author}{\bibfnamefont{A.}~\bibnamefont{Zucca}}
    (\bibinfo{year}{2021}), \eprint{2107.12990}.
  
  \bibitem[{\citenamefont{Torrado and Lewis}(2021)}]{Torrado:2020dgo}
  \bibinfo{author}{\bibfnamefont{J.}~\bibnamefont{Torrado}} \bibnamefont{and}
    \bibinfo{author}{\bibfnamefont{A.}~\bibnamefont{Lewis}},
    \bibinfo{journal}{JCAP} \textbf{\bibinfo{volume}{05}}, \bibinfo{pages}{057}
    (\bibinfo{year}{2021}), \eprint{2005.05290}.
  
  \bibitem[{cob()}]{cobaya}
  \bibinfo{howpublished}{\url{https://cobaya.readthedocs.io}}.
  
  \bibitem[{\citenamefont{Lewis}(2013)}]{Lewis:2013hha}
  \bibinfo{author}{\bibfnamefont{A.}~\bibnamefont{Lewis}},
    \bibinfo{journal}{Phys. Rev. D} \textbf{\bibinfo{volume}{87}},
    \bibinfo{pages}{103529} (\bibinfo{year}{2013}), \eprint{1304.4473}.
  
  \bibitem[{\citenamefont{Vainshtein}(1972)}]{Vainshtein:1972sx}
  \bibinfo{author}{\bibfnamefont{A.~I.} \bibnamefont{Vainshtein}},
    \bibinfo{journal}{Phys. Lett.} \textbf{\bibinfo{volume}{B39}},
    \bibinfo{pages}{393} (\bibinfo{year}{1972}).
  
  \bibitem[{\citenamefont{Damour and Polyakov}(1994)}]{Damour:1994zq}
  \bibinfo{author}{\bibfnamefont{T.}~\bibnamefont{Damour}} \bibnamefont{and}
    \bibinfo{author}{\bibfnamefont{A.~M.} \bibnamefont{Polyakov}},
    \bibinfo{journal}{Nucl. Phys.} \textbf{\bibinfo{volume}{B423}},
    \bibinfo{pages}{532} (\bibinfo{year}{1994}), \eprint{hep-th/9401069}.
  
  \bibitem[{\citenamefont{Khoury and Weltman}(2004)}]{Khoury:2003aq}
  \bibinfo{author}{\bibfnamefont{J.}~\bibnamefont{Khoury}} \bibnamefont{and}
    \bibinfo{author}{\bibfnamefont{A.}~\bibnamefont{Weltman}},
    \bibinfo{journal}{Phys. Rev. Lett.} \textbf{\bibinfo{volume}{93}},
    \bibinfo{pages}{171104} (\bibinfo{year}{2004}), \eprint{astro-ph/0309300}.
  
  \bibitem[{\citenamefont{Hinterbichler and Khoury}(2010)}]{Hinterbichler:2010es}
  \bibinfo{author}{\bibfnamefont{K.}~\bibnamefont{Hinterbichler}}
    \bibnamefont{and} \bibinfo{author}{\bibfnamefont{J.}~\bibnamefont{Khoury}},
    \bibinfo{journal}{Phys. Rev. Lett.} \textbf{\bibinfo{volume}{104}},
    \bibinfo{pages}{231301} (\bibinfo{year}{2010}), \eprint{1001.4525}.
  
  \bibitem[{\citenamefont{Bean and Tangmatitham}(2010)}]{Bean:2010zq}
  \bibinfo{author}{\bibfnamefont{R.}~\bibnamefont{Bean}} \bibnamefont{and}
    \bibinfo{author}{\bibfnamefont{M.}~\bibnamefont{Tangmatitham}},
    \bibinfo{journal}{Phys. Rev. D} \textbf{\bibinfo{volume}{81}},
    \bibinfo{pages}{083534} (\bibinfo{year}{2010}), \eprint{1002.4197}.
  
  \bibitem[{\citenamefont{Linder}(2005)}]{Linder:2005in}
  \bibinfo{author}{\bibfnamefont{E.~V.} \bibnamefont{Linder}},
    \bibinfo{journal}{Phys. Rev.} \textbf{\bibinfo{volume}{D72}},
    \bibinfo{pages}{043529} (\bibinfo{year}{2005}), \eprint{astro-ph/0507263}.
  
  \bibitem[{\citenamefont{Bertschinger and Zukin}(2008)}]{Bertschinger:2008zb}
  \bibinfo{author}{\bibfnamefont{E.}~\bibnamefont{Bertschinger}}
    \bibnamefont{and} \bibinfo{author}{\bibfnamefont{P.}~\bibnamefont{Zukin}},
    \bibinfo{journal}{Phys. Rev.} \textbf{\bibinfo{volume}{D78}},
    \bibinfo{pages}{024015} (\bibinfo{year}{2008}), \eprint{0801.2431}.
  
  \bibitem[{\citenamefont{Damour et~al.}(1990)\citenamefont{Damour, Gibbons, and
    Gundlach}}]{Damour:1990tw}
  \bibinfo{author}{\bibfnamefont{T.}~\bibnamefont{Damour}},
    \bibinfo{author}{\bibfnamefont{G.~W.} \bibnamefont{Gibbons}},
    \bibnamefont{and} \bibinfo{author}{\bibfnamefont{C.}~\bibnamefont{Gundlach}},
    \bibinfo{journal}{Phys. Rev. Lett.} \textbf{\bibinfo{volume}{64}},
    \bibinfo{pages}{123} (\bibinfo{year}{1990}).
  
  \bibitem[{\citenamefont{Brax et~al.}(2012)\citenamefont{Brax, Davis, Li, and
    Winther}}]{Brax:2012gr}
  \bibinfo{author}{\bibfnamefont{P.}~\bibnamefont{Brax}},
    \bibinfo{author}{\bibfnamefont{A.-C.} \bibnamefont{Davis}},
    \bibinfo{author}{\bibfnamefont{B.}~\bibnamefont{Li}}, \bibnamefont{and}
    \bibinfo{author}{\bibfnamefont{H.~A.} \bibnamefont{Winther}},
    \bibinfo{journal}{Phys. Rev.} \textbf{\bibinfo{volume}{D86}},
    \bibinfo{pages}{044015} (\bibinfo{year}{2012}), \eprint{1203.4812}.
  
  \bibitem[{\citenamefont{Chevallier and Polarski}(2001)}]{Chevallier:2000qy}
  \bibinfo{author}{\bibfnamefont{M.}~\bibnamefont{Chevallier}} \bibnamefont{and}
    \bibinfo{author}{\bibfnamefont{D.}~\bibnamefont{Polarski}},
    \bibinfo{journal}{Int. J. Mod. Phys. D} \textbf{\bibinfo{volume}{10}},
    \bibinfo{pages}{213} (\bibinfo{year}{2001}), \eprint{gr-qc/0009008}.
  
  \bibitem[{\citenamefont{Linder}(2003)}]{Linder:2002et}
  \bibinfo{author}{\bibfnamefont{E.~V.} \bibnamefont{Linder}},
    \bibinfo{journal}{Phys. Rev. Lett.} \textbf{\bibinfo{volume}{90}},
    \bibinfo{pages}{091301} (\bibinfo{year}{2003}), \eprint{astro-ph/0208512}.
  
  \bibitem[{\citenamefont{Weller and Lewis}(2003)}]{Weller:2003hw}
  \bibinfo{author}{\bibfnamefont{J.}~\bibnamefont{Weller}} \bibnamefont{and}
    \bibinfo{author}{\bibfnamefont{A.~M.} \bibnamefont{Lewis}},
    \bibinfo{journal}{Mon. Not. Roy. Astron. Soc.}
    \textbf{\bibinfo{volume}{346}}, \bibinfo{pages}{987} (\bibinfo{year}{2003}),
    \eprint{astro-ph/0307104}.
  
  \bibitem[{\citenamefont{Fang et~al.}(2008)\citenamefont{Fang, Hu, and
    Lewis}}]{Fang:2008sn}
  \bibinfo{author}{\bibfnamefont{W.}~\bibnamefont{Fang}},
    \bibinfo{author}{\bibfnamefont{W.}~\bibnamefont{Hu}}, \bibnamefont{and}
    \bibinfo{author}{\bibfnamefont{A.}~\bibnamefont{Lewis}},
    \bibinfo{journal}{Phys. Rev. D} \textbf{\bibinfo{volume}{78}},
    \bibinfo{pages}{087303} (\bibinfo{year}{2008}), \eprint{0808.3125}.
  
  \bibitem[{\citenamefont{Amendola et~al.}(2018)}]{Amendola:2016saw}
  \bibinfo{author}{\bibfnamefont{L.}~\bibnamefont{Amendola}}
    \bibnamefont{et~al.}, \bibinfo{journal}{Living Rev. Rel.}
    \textbf{\bibinfo{volume}{21}}, \bibinfo{pages}{2} (\bibinfo{year}{2018}),
    \eprint{1606.00180}.
  
  \bibitem[{\citenamefont{Barros et~al.}(2019)\citenamefont{Barros, Amendola,
    Barreiro, and Nunes}}]{Barros:2018efl}
  \bibinfo{author}{\bibfnamefont{B.~J.} \bibnamefont{Barros}},
    \bibinfo{author}{\bibfnamefont{L.}~\bibnamefont{Amendola}},
    \bibinfo{author}{\bibfnamefont{T.}~\bibnamefont{Barreiro}}, \bibnamefont{and}
    \bibinfo{author}{\bibfnamefont{N.~J.} \bibnamefont{Nunes}},
    \bibinfo{journal}{JCAP} \textbf{\bibinfo{volume}{01}}, \bibinfo{pages}{007}
    (\bibinfo{year}{2019}), \eprint{1802.09216}.
  
  \bibitem[{\citenamefont{Baldi}(2013)}]{Baldi:2012ua}
  \bibinfo{author}{\bibfnamefont{M.}~\bibnamefont{Baldi}}, \bibinfo{journal}{Mon.
    Not. Roy. Astron. Soc.} \textbf{\bibinfo{volume}{428}}, \bibinfo{pages}{2074}
    (\bibinfo{year}{2013}), \eprint{1206.2348}.
  
  \bibitem[{\citenamefont{Ma and Bertschinger}(1995)}]{Ma:1995ey}
  \bibinfo{author}{\bibfnamefont{C.-P.} \bibnamefont{Ma}} \bibnamefont{and}
    \bibinfo{author}{\bibfnamefont{E.}~\bibnamefont{Bertschinger}},
    \bibinfo{journal}{Astrophys. J.} \textbf{\bibinfo{volume}{455}},
    \bibinfo{pages}{7} (\bibinfo{year}{1995}), \eprint{astro-ph/9506072}.
  
  \bibitem[{\citenamefont{Hojjati et~al.}(2016)\citenamefont{Hojjati, Plahn,
    Zucca, Pogosian, Brax, Davis, and Zhao}}]{Hojjati:2015ojt}
  \bibinfo{author}{\bibfnamefont{A.}~\bibnamefont{Hojjati}},
    \bibinfo{author}{\bibfnamefont{A.}~\bibnamefont{Plahn}},
    \bibinfo{author}{\bibfnamefont{A.}~\bibnamefont{Zucca}},
    \bibinfo{author}{\bibfnamefont{L.}~\bibnamefont{Pogosian}},
    \bibinfo{author}{\bibfnamefont{P.}~\bibnamefont{Brax}},
    \bibinfo{author}{\bibfnamefont{A.-C.} \bibnamefont{Davis}}, \bibnamefont{and}
    \bibinfo{author}{\bibfnamefont{G.-B.} \bibnamefont{Zhao}},
    \bibinfo{journal}{Phys. Rev. D} \textbf{\bibinfo{volume}{93}},
    \bibinfo{pages}{043531} (\bibinfo{year}{2016}), \eprint{1511.05962}.
  
  \bibitem[{\citenamefont{Mirpoorian et~al.}(2023)\citenamefont{Mirpoorian, Wang,
    and Pogosian}}]{Mirpoorian:2023utj}
  \bibinfo{author}{\bibfnamefont{S.~H.} \bibnamefont{Mirpoorian}},
    \bibinfo{author}{\bibfnamefont{Z.}~\bibnamefont{Wang}}, \bibnamefont{and}
    \bibinfo{author}{\bibfnamefont{L.}~\bibnamefont{Pogosian}}
    (\bibinfo{year}{2023}), \eprint{2302.10999}.
  
  \bibitem[{\citenamefont{Wang et~al.}(2018)\citenamefont{Wang, Pogosian, Zhao,
    and Zucca}}]{Wang:2018fng}
  \bibinfo{author}{\bibfnamefont{Y.}~\bibnamefont{Wang}},
    \bibinfo{author}{\bibfnamefont{L.}~\bibnamefont{Pogosian}},
    \bibinfo{author}{\bibfnamefont{G.-B.} \bibnamefont{Zhao}}, \bibnamefont{and}
    \bibinfo{author}{\bibfnamefont{A.}~\bibnamefont{Zucca}}
    (\bibinfo{year}{2018}), \eprint{1807.03772}.
  
  \bibitem[{\citenamefont{Crittenden et~al.}(2009)\citenamefont{Crittenden,
    Pogosian, and Zhao}}]{Crittenden:2005wj}
  \bibinfo{author}{\bibfnamefont{R.~G.} \bibnamefont{Crittenden}},
    \bibinfo{author}{\bibfnamefont{L.}~\bibnamefont{Pogosian}}, \bibnamefont{and}
    \bibinfo{author}{\bibfnamefont{G.-B.} \bibnamefont{Zhao}},
    \bibinfo{journal}{JCAP} \textbf{\bibinfo{volume}{0912}}, \bibinfo{pages}{025}
    (\bibinfo{year}{2009}), \eprint{astro-ph/0510293}.
  
  \bibitem[{\citenamefont{Crittenden et~al.}(2012)\citenamefont{Crittenden, Zhao,
    Pogosian, Samushia, and Zhang}}]{Crittenden:2011aa}
  \bibinfo{author}{\bibfnamefont{R.~G.} \bibnamefont{Crittenden}},
    \bibinfo{author}{\bibfnamefont{G.-B.} \bibnamefont{Zhao}},
    \bibinfo{author}{\bibfnamefont{L.}~\bibnamefont{Pogosian}},
    \bibinfo{author}{\bibfnamefont{L.}~\bibnamefont{Samushia}}, \bibnamefont{and}
    \bibinfo{author}{\bibfnamefont{X.}~\bibnamefont{Zhang}},
    \bibinfo{journal}{JCAP} \textbf{\bibinfo{volume}{1202}}, \bibinfo{pages}{048}
    (\bibinfo{year}{2012}), \eprint{1112.1693}.
  
  \bibitem[{\citenamefont{Caldwell et~al.}(1998)\citenamefont{Caldwell, Dave, and
    Steinhardt}}]{Caldwell:1997ii}
  \bibinfo{author}{\bibfnamefont{R.~R.} \bibnamefont{Caldwell}},
    \bibinfo{author}{\bibfnamefont{R.}~\bibnamefont{Dave}}, \bibnamefont{and}
    \bibinfo{author}{\bibfnamefont{P.~J.} \bibnamefont{Steinhardt}},
    \bibinfo{journal}{Phys. Rev. Lett.} \textbf{\bibinfo{volume}{80}},
    \bibinfo{pages}{1582} (\bibinfo{year}{1998}), \eprint{astro-ph/9708069}.
  
  \bibitem[{\citenamefont{Aghanim et~al.}(2019)}]{Aghanim:2019ame}
  \bibinfo{author}{\bibfnamefont{N.}~\bibnamefont{Aghanim}} \bibnamefont{et~al.}
    (\bibinfo{collaboration}{Planck}) (\bibinfo{year}{2019}),
    \eprint{1907.12875}.
  
  \bibitem[{\citenamefont{Alam et~al.}(2017)}]{BOSS:2016wmc}
  \bibinfo{author}{\bibfnamefont{S.}~\bibnamefont{Alam}} \bibnamefont{et~al.}
    (\bibinfo{collaboration}{BOSS}), \bibinfo{journal}{Mon. Not. Roy. Astron.
    Soc.} \textbf{\bibinfo{volume}{470}}, \bibinfo{pages}{2617}
    (\bibinfo{year}{2017}), \eprint{1607.03155}.
  
  \bibitem[{\citenamefont{Ross et~al.}(2015)\citenamefont{Ross, Samushia,
    Howlett, Percival, Burden, and Manera}}]{Ross:2014qpa}
  \bibinfo{author}{\bibfnamefont{A.~J.} \bibnamefont{Ross}},
    \bibinfo{author}{\bibfnamefont{L.}~\bibnamefont{Samushia}},
    \bibinfo{author}{\bibfnamefont{C.}~\bibnamefont{Howlett}},
    \bibinfo{author}{\bibfnamefont{W.~J.} \bibnamefont{Percival}},
    \bibinfo{author}{\bibfnamefont{A.}~\bibnamefont{Burden}}, \bibnamefont{and}
    \bibinfo{author}{\bibfnamefont{M.}~\bibnamefont{Manera}},
    \bibinfo{journal}{Mon. Not. Roy. Astron. Soc.}
    \textbf{\bibinfo{volume}{449}}, \bibinfo{pages}{835} (\bibinfo{year}{2015}),
    \eprint{1409.3242}.
  
  \bibitem[{\citenamefont{Beutler et~al.}(2011)\citenamefont{Beutler, Blake,
    Colless, Jones, Staveley-Smith, Campbell, Parker, Saunders, and
    Watson}}]{Beutler_2011}
  \bibinfo{author}{\bibfnamefont{F.}~\bibnamefont{Beutler}},
    \bibinfo{author}{\bibfnamefont{C.}~\bibnamefont{Blake}},
    \bibinfo{author}{\bibfnamefont{M.}~\bibnamefont{Colless}},
    \bibinfo{author}{\bibfnamefont{D.~H.} \bibnamefont{Jones}},
    \bibinfo{author}{\bibfnamefont{L.}~\bibnamefont{Staveley-Smith}},
    \bibinfo{author}{\bibfnamefont{L.}~\bibnamefont{Campbell}},
    \bibinfo{author}{\bibfnamefont{Q.}~\bibnamefont{Parker}},
    \bibinfo{author}{\bibfnamefont{W.}~\bibnamefont{Saunders}}, \bibnamefont{and}
    \bibinfo{author}{\bibfnamefont{F.}~\bibnamefont{Watson}},
    \bibinfo{journal}{Monthly Notices of the Royal Astronomical Society}
    \textbf{\bibinfo{volume}{416}}, \bibinfo{pages}{3017} (\bibinfo{year}{2011}),
    \urlprefix\url{https://doi.org/10.1111%2Fj.1365-2966.2011.19250.x}.
  
  \bibitem[{\citenamefont{Abbott et~al.}(2018)}]{Abbott:2017wau}
  \bibinfo{author}{\bibfnamefont{T.~M.~C.} \bibnamefont{Abbott}}
    \bibnamefont{et~al.} (\bibinfo{collaboration}{DES}), \bibinfo{journal}{Phys.
    Rev.} \textbf{\bibinfo{volume}{D98}}, \bibinfo{pages}{043526}
    (\bibinfo{year}{2018}), \eprint{1708.01530}.
  
  \bibitem[{\citenamefont{Zhao et~al.}(2020)}]{Zhao:2020tis}
  \bibinfo{author}{\bibfnamefont{G.-B.} \bibnamefont{Zhao}} \bibnamefont{et~al.}
    (\bibinfo{year}{2020}), \eprint{2007.09011}.
  
  \bibitem[{\citenamefont{Hou et~al.}(2020)}]{Hou:2020rse}
  \bibinfo{author}{\bibfnamefont{J.}~\bibnamefont{Hou}} \bibnamefont{et~al.},
    \bibinfo{journal}{Mon. Not. Roy. Astron. Soc.}
    \textbf{\bibinfo{volume}{500}}, \bibinfo{pages}{1201} (\bibinfo{year}{2020}),
    \eprint{2007.08998}.
  
  \bibitem[{\citenamefont{Bautista et~al.}(2020)}]{Bautista:2020ahg}
  \bibinfo{author}{\bibfnamefont{J.~E.} \bibnamefont{Bautista}}
    \bibnamefont{et~al.}, \bibinfo{journal}{Mon. Not. Roy. Astron. Soc.}
    \textbf{\bibinfo{volume}{500}}, \bibinfo{pages}{736} (\bibinfo{year}{2020}),
    \eprint{2007.08993}.
  
  \bibitem[{\citenamefont{de~Mattia et~al.}(2021)}]{deMattia:2020fkb}
  \bibinfo{author}{\bibfnamefont{A.}~\bibnamefont{de~Mattia}}
    \bibnamefont{et~al.}, \bibinfo{journal}{Mon. Not. Roy. Astron. Soc.}
    \textbf{\bibinfo{volume}{501}}, \bibinfo{pages}{5616} (\bibinfo{year}{2021}),
    \eprint{2007.09008}.
  
  \bibitem[{\citenamefont{Neveux et~al.}(2020)}]{Neveux:2020voa}
  \bibinfo{author}{\bibfnamefont{R.}~\bibnamefont{Neveux}} \bibnamefont{et~al.},
    \bibinfo{journal}{Mon. Not. Roy. Astron. Soc.}
    \textbf{\bibinfo{volume}{499}}, \bibinfo{pages}{210} (\bibinfo{year}{2020}),
    \eprint{2007.08999}.
  
  \bibitem[{\citenamefont{Wang et~al.}(2020)}]{Wang:2020tje}
  \bibinfo{author}{\bibfnamefont{Y.}~\bibnamefont{Wang}} \bibnamefont{et~al.}
    (\bibinfo{year}{2020}), \eprint{2007.09010}.
  
  \bibitem[{\citenamefont{du~Mas~des Bourboux
    et~al.}(2020)}]{duMasdesBourboux:2020pck}
  \bibinfo{author}{\bibfnamefont{H.}~\bibnamefont{du~Mas~des Bourboux}}
    \bibnamefont{et~al.} (\bibinfo{year}{2020}), \eprint{2007.08995}.
  
  \bibitem[{\citenamefont{Scolnic et~al.}(2018)}]{Pan-STARRS1:2017jku}
  \bibinfo{author}{\bibfnamefont{D.~M.} \bibnamefont{Scolnic}}
    \bibnamefont{et~al.} (\bibinfo{collaboration}{Pan-STARRS1}),
    \bibinfo{journal}{Astrophys. J.} \textbf{\bibinfo{volume}{859}},
    \bibinfo{pages}{101} (\bibinfo{year}{2018}), \eprint{1710.00845}.
  
  \bibitem[{\citenamefont{Peirone et~al.}(2018)\citenamefont{Peirone, Koyama,
    Pogosian, Raveri, and Silvestri}}]{Peirone:2017ywi}
  \bibinfo{author}{\bibfnamefont{S.}~\bibnamefont{Peirone}},
    \bibinfo{author}{\bibfnamefont{K.}~\bibnamefont{Koyama}},
    \bibinfo{author}{\bibfnamefont{L.}~\bibnamefont{Pogosian}},
    \bibinfo{author}{\bibfnamefont{M.}~\bibnamefont{Raveri}}, \bibnamefont{and}
    \bibinfo{author}{\bibfnamefont{A.}~\bibnamefont{Silvestri}},
    \bibinfo{journal}{Phys. Rev.} \textbf{\bibinfo{volume}{D97}},
    \bibinfo{pages}{043519} (\bibinfo{year}{2018}), \eprint{1712.00444}.
  
  \bibitem[{\citenamefont{Bonvin and Pogosian}(2022)}]{Bonvin:2022tii}
  \bibinfo{author}{\bibfnamefont{C.}~\bibnamefont{Bonvin}} \bibnamefont{and}
    \bibinfo{author}{\bibfnamefont{L.}~\bibnamefont{Pogosian}}
    (\bibinfo{year}{2022}), \eprint{2209.03614}.
  
  \end{thebibliography}
\end{document}